\documentclass[aps, prb, twocolumn,amsmath,amssymb,floatfix, reprint]{revtex4-2}
\usepackage{graphicx}
\usepackage{svg}
\usepackage{times}
\usepackage{dcolumn}
\usepackage{hyperref}
\hypersetup{colorlinks, allcolors=blue}
\begin{document}
\title{Dominant superconducting correlations in a Luttinger liquid induced by spin fluctuations}
\author{Niels Henrik Aase}
\affiliation{\mbox{Center for Quantum Spintronics, Department of Physics, Norwegian University of Science and Technology, NO-7491 Trondheim, Norway}}
\author{Asle Sudb{\o}}
\email[Corresponding author: ]{asle.sudbo@ntnu.no}
\affiliation{\mbox{Center for Quantum Spintronics, Department of Physics, Norwegian University of Science and Technology, NO-7491 Trondheim, Norway}}

 \begin{abstract}
We study spin-fluctuation mediated divergent superconducting fluctuations in a Luttinger liquid  proximity-coupled to a spin chain.
Our study provides insight into how spin fluctuations can induce superconductivity in a strongly correlated non-Fermi liquid with repulsive electronic interactions only. The electrons in the system are governed by the Extended Hubbard Hamiltonian and are coupled to a chain of localized spins modeled by the spin-$\frac{1}{2}$ $XX$ Hamiltonian. Using a multichannel Luttinger liquid approach, we determine the phase diagram of the metal chain. We find that spin-polarized triplet superconducting correlations persist for repulsive electronic interactions for sufficiently large interchain couplings.
\end{abstract}

\maketitle

\textit{Introduction.} 
During the last decade, experimental progress in nanoengineering has allowed for unprecedented control over structures with pronounced physical properties in the quantum domain \cite{Geim2013, Aspelmeyer2014, Clerk2020, Torre2021}. Many of the nanostructures are interesting in their own right, as several long-standing predictions have been probed directly \cite{Bockrath1999, Jotzu2014, Toskovic2016}. However, with the ability to use nanostructures as fundamental building blocks, one can also construct complex heterostructure where the emergent physics is richer than the sum of its constituent parts. The advent of a significant array of experimentally realizable low-dimensional structures, motivates renewed efforts on the theoretical side to determine new avenues to be pursued. 

One such avenue is low-dimensional hybrid structures involving gapless fermionic surface states, magnetic insulators, and superconductors. These systems have received much attention already both theoretically \cite{Sau2010, Amundsen2018, Rex2020, Diaz2021} and experimentally \cite{Gibertini2019, Liu2020, Deng2020}, especially over the last decade, following the discovery of metallic surface states in topological insulators \cite{Konig2007, Hsieh2008, Hasan2010}.
In search of Majorana fermions, suggested as fundamental building blocks in topological quantum computers, one-dimensional hybrid structures in particular have been the subject of intense investigations \cite{Kitaev2001, Mourik2012, Albrecht2016, Liu2017, Lutchyn2018, Zhang2019, Vaitiekenas2021, Flensberg2021, Pan2022, Escribano2022}.
For two-dimensional systems, heterostructures of this type have been investigated in the context of obtaining spin-polarized supercurrents with potential applications to superconducting spintronics \cite{Linder2015, Eschrig2015}.
Furthermore, planar interfaces consisting of, on the one hand, metals or metallic surface states of topological insulators, and ferromagnetic or antiferromagnetic insulators, on the other hand, have been studied in the context of magnon-mediated unconventional superconductors \cite{Kargarian2016, Hugdal2018, Rohling2018,Fjaerbu2019, Erlandsen2019,Erlandsen2020, Thingstad2021}.

In many studies on low-dimensional hybrid systems, the metallic states are mostly modeled using a well-defined single-particle physics picture for the electrons that are proximity coupled to other states in the heterostructure.
On the other hand, it is well known that in one-dimensional systems, any amount of two-body scattering suffices to destroy the one-to-one correspondence between the interacting and noninteracting low-energy excitations.
The resulting fixed point, the Luttinger liquid \cite{Tomonaga1950, Luttinger1963}, is one where low-energy fermionic excitations of the noninteracting case are replaced by well-defined bosons \cite{Haldane1981}, describing collective density fluctuations in the spin and charge sectors.
In the context of magnon-mediated unconventional superconductivity in low-dimensional heterostructures, it is thus of some interest to consider the fate of the superconducting state when it no longer arises out of a Fermi liquid.
Similar issues need to be considered in the context of high-$T_c$ superconductivity in cuprate oxides \cite{Anderson1997, Lederer2017}.
In this paper, we therefore revisit the question of if and how superconductivity arises when a one-dimensional interacting fermion chain with gapless fermions interacts with a one-dimensional chain of localized spins. We employ simple lattice models for both components of the hybrid structure, and treat them using a multichannel Luttinger liquid approach \cite{Sandler1997, Mukhopadhyay2001, Yurkevich2013, Kagalovsky2017, Jones2017, Yurkevich2017}.

\textit{Microscopic model.}
To model the fermion chain, we use the extended Hubbard (EHB) Hamiltonian, $H_{\mathrm{EHB}}$, which has been extensively studied in one dimension \cite{Voit1992,Tsuchiizu2002, Menard2011, Ejima2007, Iemini2015, Spalding2019}. In terms of annihilation and creation operators  $c_{i\sigma}^\dagger$ and $c_{i\sigma}$ for electrons on site $i$ with spin $\sigma$, $H_{\mathrm{EHB}}$ can be expressed as
\begin{align}
\label{H_EHB}
\begin{split}
        H_{\mathrm{EHB}} &= -t\sum_{i, \sigma} c_{i\sigma}^\dagger c_{i+1, \sigma} -\mu\sum_i n_i \\
        &+ U\sum_i n_{i\uparrow}n_{i\downarrow} + V\sum_{i}n_{i}n_{i + 1} ,
\end{split}
\end{align}
where $n_{i\sigma} = c_{i\sigma}^\dagger c_{i\sigma}$, $n_i = n_{i\uparrow} + n_{i\downarrow}$, $t$ is the hopping amplitude between adjacent sites, $\mu$ is the chemical potential, $U$ is the onsite interaction, and $V$ is the interaction between electrons situated on adjacent sites.
The quantum spin operators $\boldsymbol{S}_i = (S_i^x, S_i^y, S_i^z)$, satisfying the commutation relation $[S_i^\alpha, S_i^\beta] = i\hbar\epsilon_{\alpha\beta\gamma} S_i^\gamma$, are used to describe the spin chain, modeled by the spin-$\frac{1}{2}$ $XX$ Hamiltonian, $H_{\mathrm{FMI}}$
\begin{equation}
    H_{\mathrm{FMI}} = -J_{xy} \sum_{i}(S^x_{i}S^x_{i+1} + S^y_{i}S^y_{i+1}),
    \label{H_FMI}
\end{equation}
where $J_{xy}$ is the ferromagnetic exchange coupling. 
The interchain coupling, denoted $H_{\mathrm{int}}$, is parametrized by $\bar{J}$, and is given by
\begin{equation}
\label{H_int}
    H_{\mathrm{int}} = -2\bar{J}\sum_i (c_{i\uparrow}^\dagger, c_{i\downarrow}^\dagger) \boldsymbol{\tau} (c_{i\uparrow}, c_{i\downarrow})^{\mathrm{T}} \cdot \boldsymbol{S}_i,
\end{equation}
inspired by \cite{Erlandsen2020}, where $\boldsymbol{\tau}$ is a vector of the Pauli matrices, acting on the electron spin degree of freedom.
The Hamiltonian for the entire system is $H = H_{\mathrm{EHB}} + H_{\mathrm{FMI}} +H_{\mathrm{int}}$ and thus describes a one-dimensional Kondo lattice with additional electron-electron and spin-spin interactions. Similar systems have been considered within the Luttinger liquid framework  \cite{Maciejko2012, Altshuler2013, Yevtushenko2015, Tsvelik2019} and can be realized by coupling helical edge states in topological insulators to spin impurities.
Note that because $J_z$ is zero, $H$ is not $SU(2)$ symmetric. For convenience, we employ natural units $\hbar =1$ and use $t$ as the unit of energy in $H$.

For our purposes, the most suitable approach to the problem is to employ the Jordan-Wigner transformation \cite{Jordan1928}, as it allows a unified treatment of both chains. By introducing $S^\pm_i = (S_i^x \pm iS_i^y)$, and the spinless fermion operators $d_i^\dagger$, $d_i$, the well-known mappings $S_i^\pm = d_i^{(\dagger)}\mathrm{e}^{\pm i\pi\sum_{n=1}^{i-1} d_n^\dagger d_n}$ and $S_i^z = d_i^\dagger d_i - \frac{1}{2}$ are established. By extracting the cubic terms from $H_{\mathrm{int}} = H_{\mathrm{int}}^{\mathrm{c}} + H'_{\mathrm{int}}$, the string operator vanishes when inserting the fermion operators into $H'_{\mathrm{int}}$ and $H_{\mathrm{FMI}}$
\begin{align}
    H_{\mathrm{FMI}} &= -J_{xy} \sum_{i}d^\dagger_{i}d_{i+1} \label{H_FMI_fermion} \\
    H'_{\mathrm{int}} &= -2\bar{J}\sum_i (c_{i\uparrow}^\dagger c_{i\uparrow} - c_{i\downarrow}^\dagger c_{i\downarrow})(d^\dagger_id_i -\frac{1}{2}) \label{H_int_fermions}.
\end{align}
From equation \eqref{H_int_fermions}, it follows that $\bar{J}$ acts as both the strength of the chain coupling and as an effective magnetic field in the $z$ direction felt only by the metal chain. The latter is accounted for by introducing a spin dependency in the chemical potential $\mu_\sigma = \mu - \sigma \bar{J}$. Equation \eqref{H_FMI_fermion} shows that $J_{xy}$ plays the role of a hopping parameter in the spin chain. $H_{\mathrm{int}}^\mathrm{c}$ will be discussed further in the next section.

All three species of fermions have the same kinetic structure. Their dispersion relations are $\varepsilon_l(k) = -2t_l\cos(k) -\mu_l$, with $l$ being the species index $l = (\uparrow, \downarrow, S)$. From this, one finds the Fermi momentum and Fermi velocity, $k_{\mathrm{F}}^l = \arccos(-\mu_l/(2t_l))$ and $v_{\mathrm{F}}^l = 2t_l\sin(k_{\mathrm{F}}^l)$, respectively. The spin chain has $v_{\mathrm{F}}^\mathrm{S} = 2J_{xy}$ and $k_{\mathrm{F}}^\mathrm{S}=\pi/2$, physically corresponding to the absence of any net magnetization in the $z$ direction arising due to terms in $H_{\mathrm{FMI}}$. $k_{\mathrm{F}}^\sigma$ is dependent on $\mu_\sigma$, such that $\mu$ determines $k_{\mathrm{F}}$ in the absence of any chain coupling, while $\bar{J}$ controls the extent of the spin splitting. 

\textit{Continuum limit field theory.}
To describe the low-energy physics of our system, we use bosonization \cite{Delft1998}. 
The low-energy excitations are described by linearizing the spectrum of the noninteracting case around the two Fermi points $\pm k^l_{\mathrm{F}}$. Annihilation operators can then be written as $c_{il} = \sum_s \psi_{sl}(x=ia)$ where $\psi_{sl}(x)$ destroys a fermion of species $l$ on the branch $s=\pm$. By extending the linearized spectrum to $\pm \infty$, using a soft cutoff, and taking the continuum limit, the following operator identity holds \cite{Haldane1981} 
\begin{equation}
        \psi_{sl}(x) =  \lim_{\alpha \to 0} \frac{U_{sl} }{\sqrt{2\pi\alpha}} \mathrm{e}^{ir(k_{\mathrm{F}}^l - \pi/L)x} 
        \mathrm{e}^{-i(s\phi_l(x) -\theta_l(x))}.
        \label{boson_identity}
\end{equation}
Here, $U_{sl}$ is a Klein factor which has the effect of ensuring correct fermionic anticommutation relations and moreover of raising or lowering the number of fermions in the system \cite{Delft1998}, $\alpha$ is a cutoff ensuring finite bandwidth, and $\phi$ and $\theta$ are bosonic fields. The details of the construction of $\phi$ and $\theta$ and their explicit representation can be found in several reviews on abelian bosonization \cite{Voit1995, Fradkin2013,Giamarchi2003}, and will not be repeated here. Due to the relations $\nabla \phi(x) = -\pi(n_R(x) + n_L(x))$ and $\nabla \theta(x) = -\pi(n_R(x) - n_L(x))$, $\phi$ and $\theta$ can be interpreted as density and current fields, respectively.

For repulsive $U$, using renormalization-group theory one finds that the backscattering term is irrelevant. The low-energy physics of the model in the presence of a magnetic field is then described by the Tomonaga-Luttinger (TL) model \cite{Penc1993}. For $U<0$, the backscattering term is {\it a priori} relevant and gaps the spin sector. In the presence of a sufficiently strong magnetic field, backscattering is however suppressed. This readmits a TL representation \cite{Zhao2008}. The absence of large momentum transfers can be attributed to the Fermi momentum mismatch between opposite-spin electrons $\delta k_{\mathrm{F}} = k_{\mathrm{F}}^\downarrow -k_{\mathrm{F}}^\uparrow$
with $\bar{J}$ acting as an effective magnetic field. The same will hold for the system we consider, especially since we will focus on the parameter regime where $\bar{J}$, and thus also $\delta k_{\mathrm{F}}$, is large. 
Furthermore, we use bosonic fields  $\phi_\mathrm{S}$ and $\theta_\mathrm{S}$ associated to the spin chain to represent $S_i^\pm$ \cite{Giamarchi2003}. It then follows that the terms in $H_{\mathrm{int}}^\mathrm{c}$ are a product of two complex exponentials. The first is a linear combination of slowly varying fields, while the other is $\mathrm{e}^{i(\delta k_\mathrm{F} + n k_\mathrm{F}^\mathrm{S})x}$ with $n=0,2$. Since $\delta k_{\mathrm{F}}>0$, the latter exponential oscillates rapidly. Thus, when integrating over the length of the system, the cubic terms average to zero and may be neglected.

By the preceding argument, it follows that only terms quadratic in the fields remain in interactions between different fermion species, as they have different Fermi momenta.
The same-spin interactions between nearest neighbors require more care. In the weak-coupling regime they take the form \cite{Capponi2000}
\begin{equation}
    V\sum_i n_{i+1, \lambda}n_{i,\lambda} = \int \mathrm{d} x \frac{1}{\pi^2}V(1-\cos(2k_{\mathrm{F}}^\lambda a))(\nabla\phi_\lambda(x))^2,
    \label{V_same_spin_fields}
\end{equation}
where $\lambda = \uparrow,\downarrow$. 
In general we will avoid half-filling, and any accidental Umklapp scattering in the metal chain arising if either spin band is at half-filling is neglected. Because $J_z=0$, there is no Umklapp scattering in the spin chain either, yielding a purely quadratic theory describing a TL liquid.

We next introduce in standard fashion bosonic fields associated with the charge and spin densities in the TL liquid originating with $H_{\rm{EHB}}$, $n_{i\rho} = (n_{i\uparrow} + n_{i\downarrow})/\sqrt{2}$ and $n_{i\sigma} = (n_{i\uparrow} - n_{i\downarrow})/\sqrt{2}$, respectively. This will also accentuate the magnetic nature of the interchain coupling in equation \eqref{H_int_fermions}. Employing the bases $\boldsymbol{\phi} = (\phi_\rho, \phi_\sigma,\phi_\mathrm{S})^{\mathrm{T}}$ describing densities in the three channels $(\rho, \sigma,S)$ and $\boldsymbol{\theta} = (\theta_\rho, \theta_\sigma, \theta_\mathrm{S})^{\mathrm{T}}$ describing currents in the same three channels, one obtains from \eqref{V_same_spin_fields} the following expression for $H$:
\begin{equation}
    H = \frac{1}{2\pi} \int \mathrm{d}x \partial_x\begin{pmatrix}
        \boldsymbol{\phi}^{\mathrm{T}} & \boldsymbol{\theta}^{\mathrm{T}}
    \end{pmatrix}
    \begin{pmatrix}
            V_\phi & 0 \\
            0 & V_\theta
        \end{pmatrix} \partial_x\begin{pmatrix}
        \boldsymbol{\phi} \\ \boldsymbol{\theta}
    \end{pmatrix}. 
    \label{H_fields}
\end{equation}
The symmetric matrices $V_\phi$ and $V_\theta$ contain all microscopic details of the model. $V_\theta$ is 
\begin{equation}
    V_\theta = \begin{pmatrix}
        \bar{v_{\mathrm{F}}} & \delta v_{\mathrm{F}} & 0 \\
        \delta v_{\mathrm{F}} & \bar{v_{\mathrm{F}}} & 0 \\
        0 & 0 & v_{\mathrm{F}}^\mathrm{S}
    \end{pmatrix}, \label{V_theta_micro}
\end{equation}
where $\bar{v_{\mathrm{F}}} \equiv (v_{\mathrm{F}}^\uparrow + v_{\mathrm{F}}^\downarrow)/2$ and $\delta v_{\mathrm{F}} \equiv (v_{\mathrm{F}}^\uparrow - v_{\mathrm{F}}^\downarrow)/2$. The expression for $V_\phi$ is more complicated
\begin{equation}
    V_\phi = \begin{pmatrix}
        \bar{v_{\mathrm{F}}} + U_\rho/\pi & \delta v_{\mathrm{F}} + \delta V/\pi & 0 \\
        \delta v_{\mathrm{F}} + \delta V/\pi & \bar{v_{\mathrm{F}}} - U_\sigma/\pi & \frac{2\sqrt{2}\bar{J}}{\pi} \\
        0 & \frac{2\sqrt{2}\bar{J}}{\pi} & v_{\mathrm{F}}^\mathrm{S}
    \end{pmatrix}, \label{V_phi_micro}
\end{equation}
with
\begin{subequations}
\label{EHB_params}
\begin{align}
    U_\rho &= U + 4V - V(\cos(2k_{\mathrm{F}}^\uparrow a) + \cos(2k_{\mathrm{F}}^\downarrow a)) \\
    U_\sigma &= U + V(\cos(2k_{\mathrm{F}}^\uparrow a) + \cos(2k_{\mathrm{F}}^\downarrow a)) \\
    \delta V &= V(\cos(2k_{\mathrm{F}}^\downarrow a) - \cos(2k_{\mathrm{F}}^\uparrow a)).
\end{align}
\end{subequations}
From equations \eqref{V_theta_micro} and \eqref{V_phi_micro}, the influence of the intrachain coupling is seen to be twofold. Firstly, the effective magnetic field $\bar{J}$ destroys the spin-charge separation normally present in the EHB model, since the coupling between the electron spin and charge channels is nonzero. Secondly, $\bar{J}$ also acts as an interchannel coupling between the electron spin channel and the channel describing the spin chain.

\textit{Multichannel Luttinger liquids and correlation functions.}
From the relation $[\phi_l(x_1), \partial_x\theta_m(x_2)] = i\delta_{lm}\delta(x_1-x_2)/\pi$ and equation \eqref{H_fields}, the action of the system is obtained:
\begin{align}
    \label{action_fields}
\begin{split}
    S[\boldsymbol{\phi}, \boldsymbol{\theta}] &= \frac{1}{2\pi} \int \mathrm{d} x \mathrm{d} \tau \begin{pmatrix}
        \boldsymbol{\phi}^{\mathrm{T}} & \boldsymbol{\theta}^{\mathrm{T}}
    \end{pmatrix}\\
    &\left[
        \begin{pmatrix}
            0 & \mathbb{I}_3 \\
            \mathbb{I}_3 & 0
        \end{pmatrix}i\partial_\tau + \begin{pmatrix}
            V_\phi & 0 \\
            0 & V_\theta
        \end{pmatrix}\partial_x
    \right] \partial_x\begin{pmatrix}
        \boldsymbol{\phi} \\ \boldsymbol{\theta}
    \end{pmatrix}, 
    \end{split}
\end{align}
where $\tau$ is imaginary time, $\mathbb{I}$ is the identity matrix, and the differential operators inside the square bracket act to the left. 

Equation \eqref{action_fields} describes the action of a multichannel Luttinger liquid. Such systems are often considered when introducing disorder to systems consisting of coupled quantum wires \cite{Sandler1997, Mukhopadhyay2001}.
 We emphasize that  our system differs from these, in that we employ the multichannel Luttinger liquid formalism to one-dimensional systems consisting of both electrons and localized magnetic moments.
In the setting of coupled quantum wires, a method for mapping the case of interchannel interactions, back to the well known case of diagonal interaction matrices has been devised \cite{Yurkevich2013, Kagalovsky2017, Jones2017}. Introducing the matrix $M$ with the properties $M^{\mathrm{T}}V_\phi M = M^{-1} V_\theta M^{-\mathrm{T}} = u$, where $u$ is a diagonal matrix congruent to both $V_\phi$ and $V_\theta$. Introducing the transformed fields, $\tilde{\boldsymbol{\phi}} = M^{-1}\boldsymbol{\phi}$ and $\tilde{\boldsymbol{\theta}} = M^{\mathrm{T}}\boldsymbol{\theta}$, the first term in equation \eqref{action_fields} is left invariant, while the second term is diagonalized:
\begin{align}
    \label{action_fields_diag}
\begin{split}
    S[\tilde{\boldsymbol{\phi}}, \tilde{\boldsymbol{\theta}}] &= \frac{1}{2\pi} \int  \mathrm{d} x \mathrm{d} \tau \begin{pmatrix}
        \tilde{\boldsymbol{\phi}}^{\mathrm{T}} & \tilde{\boldsymbol{\theta}}^{\mathrm{T}}
    \end{pmatrix} \\
    &\left[
        \begin{pmatrix}
            0 & \mathbb{I}_3 \\
            \mathbb{I}_3 & 0
        \end{pmatrix}i\partial_\tau + \begin{pmatrix}
            u & 0 \\
            0 & u
        \end{pmatrix}\partial_x
    \right] \partial_x\begin{pmatrix}
        \tilde{\boldsymbol{\phi}} \\ \tilde{\boldsymbol{\theta}}
    \end{pmatrix}.
\end{split}
\end{align}
$M$ is constructed using the procedure presented in Ref.\ \cite{Yurkevich2017}. Similar approaches are used in Refs.\ \cite{Affleck2013, Jones2017PHD} for the two-channel case. The entries in $u$ are the velocities of the three types of collective excitations in the system. Since the relation between $(\boldsymbol{\phi}, \boldsymbol{\theta})$ and $(\tilde{\boldsymbol{\phi}}, \tilde{\boldsymbol{\theta}})$ is known, calculating correlation functions is effectuated by a change of basis and using equation \eqref{action_fields_diag}. To this end, we introduce the symmetric Luttinger matrix $K = M M^{\mathrm{T}}$, which will play a role corresponding to the Luttinger parameter for the single-channel case, i.e., the entries are determined by the parameters of the model.

By using equation \eqref{action_fields_diag}, it is straightforward to calculate correlation functions in the form
\begin{equation}
    I = \langle  \mathrm{exp}(i\sum_i \boldsymbol{A}_i^{\mathrm{T}} \boldsymbol{\phi}(r_i) + \boldsymbol{B}_i^{\mathrm{T}} \boldsymbol{\theta}(r_i)) \rangle.
    \label{I_integral}
\end{equation}
We refer to the Supplemental Material for details \cite{Suppl}. Here, we have introduced the shorthand notation $r=(x, \tau)$, and the vector components $A_i^l$ and $B_i^l$ are associated to $\boldsymbol{\phi}_l(r_i)$ and $\boldsymbol{\theta}_l(r_i)$, respectively. Assuming that $\tau_i=\tau \; \forall \; i$, the correlation function $I$ in \eqref{I_integral} is computed using the same techniques as in the single-channel case outlined in Ref.\ \cite{Giamarchi2003}. It is found that $I$ is only nonzero when $\sum_l M_{lm}\sum_iA^l_i=0$ and $\sum_l M_{lm}^{-\mathrm{T}}\sum_iB^l_i=0$. For $\boldsymbol{A}$ and $\boldsymbol{B}$ fulfilling this criterion, the expression for $I$ is
\begin{equation}
    I = \left(\frac{\alpha^2}{x^2+ \alpha^2}\right)^{-\frac{1}{4}\sum_{l,l'}\sum_{i<j}(A_i^lA_j^{l'} K_{ll'}+ B_i^lB_j^{l'} K_{ll'}^{-1})}.
    \label{I_integral_solved}
\end{equation}
The nonuniversal power-law decay, where the exponent is dependent on the microscopic details contained in $K$, is a hallmark of correlation functions in Luttinger liquids, and will be used to determine the phase diagram of the electrons in the system.

Due to the low dimensionality of the system, true long-range order is precluded even at zero temperature, but signature remnants of long-range orders can nonetheless be investigated. Choosing an order parameter (OP) $O_\eta(x,\tau)$, with $\eta$ denoting the type of order, the associated correlation function $R_\eta(x) = \langle O_\eta(x, 0)O_\eta^\dagger(0,0) \rangle \propto x^{-\nu_\eta}$ may be studied, where $\nu_\eta$ are nonuniversal exponents. The OP with the smallest $\nu_\eta$ at zero temperature, corresponding to the most strongly divergent susceptibility, will then identify the phase.
The phase diagram of the electrons in the metal chain is thus determined by the interactions present in the system in that they determine the various $\nu_\eta$. 

For repulsive interactions, it becomes favorable for the electrons in the metal chain to enter either a charge density wave (CDW) or a spin density wave (SDW) phase. For attractive interactions, the electrons pair up in either a singlet state (SS), or a triplet state (TS). The different OPs are \cite{Giamarchi2003}
\begin{subequations}
    \label{order_param_expressions_with_changed_spin_expressions}
    \begin{align}
        O_{\mathrm{CDW}/\mathrm{SDW}^z}(r) &= \psi_{R\uparrow}^\dagger(r)\psi_{L\uparrow}(r) \pm \psi_{R\downarrow}^\dagger(r)\psi_{L\downarrow}(r) \\
        O_{\mathrm{SDW}^{x/y}}(r) &= \psi_{R\uparrow}^\dagger(r)\psi_{L\downarrow}(r) \pm \psi_{R\downarrow}^\dagger(r)\psi_{L\uparrow}(r) \\ 
        O_{\mathrm{TS}^{\uparrow/\downarrow}}(r) &= \psi_{R\uparrow/\downarrow}^\dagger(r)\psi_{L\uparrow/\downarrow}^\dagger(r)\\
        O_{\mathrm{SS}/\mathrm{TS}^0}(r) &= \psi_{R\uparrow}^\dagger(r)\psi_{L\downarrow}^\dagger(r) \pm \psi_{L\uparrow}^\dagger(r)\psi_{R\downarrow}^\dagger(r),
    \end{align}
    \end{subequations}
where the upper (lower) sign applies to the first (latter) OP in each expression. 
\begin{figure*}[t]
    \centering
    \includegraphics[width=\textwidth,keepaspectratio]{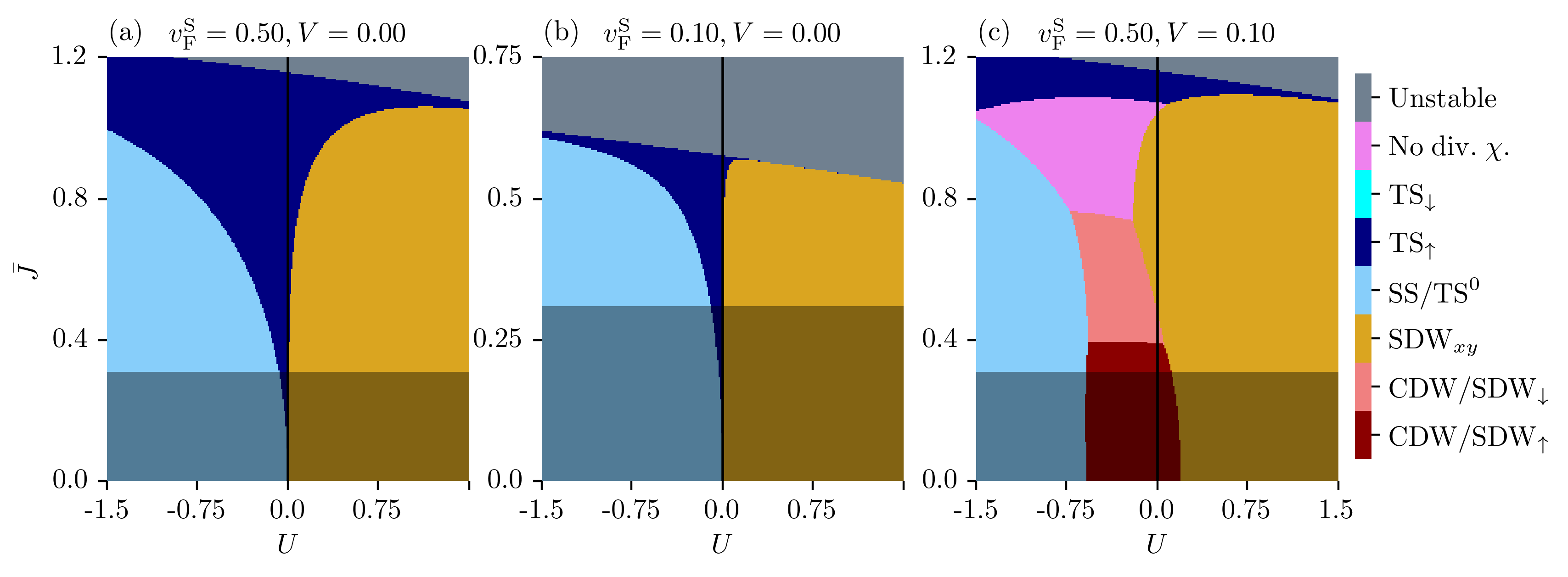}
    \caption{Phase diagrams in terms of the onsite interaction $U$ and the interchain coupling $\bar{J}$. The colors indicate the order parameter which decays the slowest; see main text for a description of the different order parameters. The Fermi velocity in the spin chain $v_\mathrm{F}^\mathrm{S}$ and the nearest-neighbor interaction $V$ vary in the three subfigures, while the Fermi momentum $k_{\mathrm{F}}a =0.45\pi$ is the same. The shaded area marks the region where $\delta k_{\mathrm{F}} a \leq 0.2$, where the quadratic theory is invalid; see main text.}
    \label{fig:phase_diag}
\end{figure*}
Inserting the single-particle expression in equation \eqref{boson_identity} into equation \eqref{order_param_expressions_with_changed_spin_expressions}, and by using equation \eqref{I_integral_solved}, the expressions for $R_\eta$ are found to be 
    \begin{subequations}
    \label{R_expressions}
    \begin{align}
        R_{{\mathrm{CDW}/\mathrm{SDW}^z}} &\propto \frac{\mathrm{e}^{2ik_{\mathrm{F}}^\uparrow x}}{x^{K_{11} + K_{22} + K_{12}}} + \frac{\mathrm{e}^{2ik_{\mathrm{F}}^\downarrow x}}{x^{K_{11} + K_{22} - K_{12}}} \label{R_CDW} \\
        R_{\mathrm{SDW}^{x/y}} &\propto \mathrm{e}^{i(k_{\mathrm{F}}^\uparrow + k_{\mathrm{F}}^\downarrow)x} \frac{1}{x^{K_{11} + K^{-1}_{22}}} \\
        R_{\mathrm{TS}^{\uparrow/\downarrow}} &\propto \frac{1}{x^{K^{-1}_{11} + K^{-1}_{22} \pm K^{-1}_{12}}} \\
        R_{\mathrm{SS}/\mathrm{TS}^0} &\propto \frac{e^{i\delta k_{\mathrm{F}}}}{x^{K^{-1}_{11} + K_{22}}} + \frac{e^{-i\delta k_{\mathrm{F}}}}{x^{K^{-1}_{11} + K_{22}}}.
    \end{align}
\end{subequations}
Note that although we are studying a three-channel system, the above correlation function exponents are given exclusively in terms of $K_{11}, K_{12}$ and $K_{22}$. 
In general, the two OPs in each expression $R_\eta$ cannot be distinguished, with the polarized TS being the exception as one can use the appropriate sign in front of $K_{12}$. This is easily understood for $R_{\mathrm{SS}/\mathrm{TS}^0}$. With $\bar{J}=0$, $V=0$, and $U<0$ the system is gapped in the spin sector, and the dominant phase is a SS. With an effective magnetic field, the SS is converted into a FFLO state \cite{Fulde1964, Larkin1964} with center of mass momentum $\pm \delta k_{\mathrm{F}}$ \cite{Yang2001,Zhao2008,Feigun2009}. In $R_{{\mathrm{CDW}/\mathrm{SDW}^z}}$ one can observe two distinct density waves, with wave numbers $2k_{\mathrm{F}}^\uparrow$ and $2k_{\mathrm{F}}^\downarrow$, each wave carrying net spin and charge. Finally, we note that the density-density correlations have an additional $k_{\mathrm{F}}$-independent term, which always exhibits Fermi liquid decay with $x^{-2}$, independent of the microscopic details.

\textit{Results and discussion.} Figure \ref{fig:phase_diag} presents three phase diagrams for our model. The dominant phase is found by calculating $K$ for every set of microscopic parameters. The smallest $\nu_\eta$ is subsequently determined using equation \eqref{R_expressions}. This identifies the dominant divergence and the most favorable phase. The colors of the figures are associated with different OPs. No divergent $\chi$ indicates that the Fermi liquid decay in the density correlations dominates. The unstable region indicates that one of the velocities in $u$ is imaginary. This may be indicative of a phase transition \cite{Kitazawa2003, Affleck2013}, sometimes referred to as a Wentzel-Bardeen (WB) singularity \cite{Wentzel1951, Bardeen1951}. When $\bar{J}$ becomes the dominant interaction, it is possible that the WB singularity arises because the system becomes phase separated, as both chains are separated into regions with equal polarization, similar to the $t$-$J$ model \cite{Ogata1991_2}. The lightly shaded region marks the area where $\delta k_{\mathrm{F}}a$ is not large enough to safely discard large momentum transfer terms. We choose the value $\delta k_{\mathrm{F}}a =0.2$ to bound this region. Since changing the sign of $\bar{J}$ is equivalent to flipping the quantization axis, all OPs insensitive to this operation are symmetric with $\bar{J}$, while spin-polarized OPs are mapped to their spin-flipped counterpart. 

In Figures \ref{fig:phase_diag} (a) and (b), $V=0$, hence the system is similar to the Hubbard models studied in Refs.\ \cite{Penc1993, Zhao2008}, with an additional channel due to the spin chain, resulting in a richer phase diagram. Despite having a different Fermi velocity in the spin chain, the two systems exhibit the same qualitative traits for attractive $U$, with the preferred state being the FFLO state. However, once $|\bar{J}|$ becomes large enough, the polarized TS is preferred, with the sign of $\bar{J}$ determining the polarization of the state. Note that this occurs in regions where the quadratic theory is valid. This transition can be understood as competing electron pairing mechanisms. An attractive $U$ favors onsite pairing with opposite spin, but this pairing is suboptimal when including interchain interactions. For large $\bar{J}$, the optimal placing of the electrons is such that they are always adjacent to a localized spin with the same spin-polarization, avoiding double occupancy of a site.

The competing interactions can be further understood by considering the properties of the Luttinger matrix $K$. While the explicit expression for $K$ in terms of the underlying microscopic parameters rapidly becomes intractable as the number of channels increases, the two-channel case has been studied in detail \cite{Jones2017,Affleck2013} and provides valuable insight into the physics of the present three-channel case. An important property for two-channel systems is that density (current) inter-channel interactions are found to enhance (suppress) the diagonal elements in $K$ \cite{Kagalovsky2017}.
In the present three-channel case, the charge channel does not couple to the spin chain, so to a first approximation, we may consider the electron spin-charge block and the spin-spin block as two distinct two-channel systems, and use the insight derived for the two-channel case \cite{Jones2017PHD} on each system separately. We emphasize that our results for $K_{ij}$ are obtained using the full interaction matrices $V_\theta$ and $V_\phi$.
The spin-charge block describes a metal chain subject to a magnetic field, and the density and current interchannel interactions neutralize each other. However, in the spin-spin block, since $V_\theta^{23} =0$, the density interaction in $V_\phi^{23}$ enhances $K_{22}$ (and $K_{33}$) in a manner which is not canceled. For some value of $\bar{J}$, the increase of $K_{22}$ causes the TS to be favored over the FFLO state in Figures \ref{fig:phase_diag} (a) and (b) for $U<0$. The sign of $K_{12}^{-1}$ determines the spin polarization of the TS. 

The size of the envelope enclosing the stable part of the phase diagrams in Figure \ref{fig:phase_diag} increases with $v_{\mathrm{F}}^{\mathrm{S}}$. This can be attributed to one of the velocities in $u$ turning imaginary. The imaginary velocity is associated with a component of $\tilde{\boldsymbol{\phi}}$ and $\tilde{\boldsymbol{\theta}}$ mainly comprised of the spin chain fermions. So for large $\bar{J}$, near the unstable region, the spin chain is dominated by the interchain interaction, as expected when $\bar{J}\gg J_{xy}$. 
While $J_{xy}$ is typically orders of magnitudes lower than the hopping parameter, the systems in Figures \ref{fig:phase_diag} (a) and (c) can be mapped to more realistic ranges of parameter values. Due to the congruence relation $K V_\phi K = V_\theta$, scaling all entries in the interaction matrices by a multiplicative factor yields the same $K$ and thus the same phase diagram. Considering a narrow bandwidth model or a sparsely populated system with small $k_\mathrm{F}$ would thus yield the same phase diagrams in Figure \ref{fig:phase_diag}, with more realistic parameters.

\begin{figure}
    \centering
    \includegraphics[width=\linewidth,keepaspectratio]{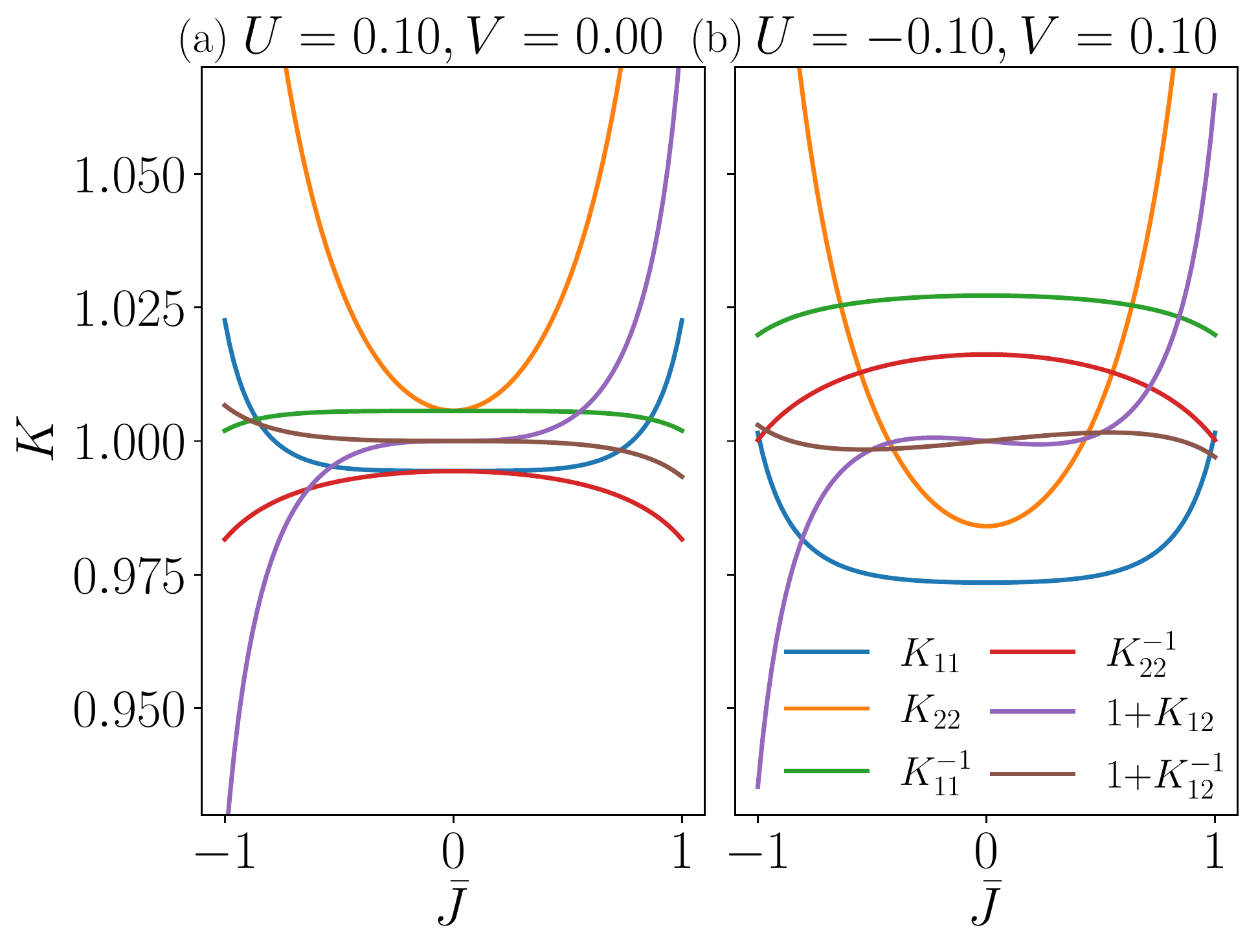}
    \caption{Elements of the Luttinger matrix $K$ as a function of the interchain coupling $\bar{J}$. The interaction strengths $U$ and $V$ are different in the two plots, while the parameters $v_\mathrm{F}^\mathrm{S}=0.5$ and $k_\mathrm{F}a/\pi=0.45$ are the same. Only the matrix relevant for the electron correlation function exponents are considered.}
    \label{fig:K_matrix_elements}
\end{figure}

The $\mathrm{SDW}_{xy}$ correlations are dominant in the repulsive sector of Figure \ref{fig:phase_diag} (b), except for $U\ll \bar{J}$. This area of the phase diagram is expanded as the envelope size increases with $v_{\mathrm{F}}^{\mathrm{S}}$ in Figure \ref{fig:phase_diag} (a), where the TS state is dominant also for larger values of $\bar{J}/U$. To elucidate how this occurs, we go beyond the analytical results for the two-channel case and plot the various matrix elements of $K$ as a function of $\bar{J}$ in Figure \ref{fig:K_matrix_elements}. Some properties are independent of other parameters: The (off-)diagonal elements of both $K$ and $K^{-1}$ are (anti-)symmetric in $\bar{J}$, $K$ ($K^{-1}$ ) increases (decreases) with $\bar{J}$, and the spin channel is more strongly dependent on $\bar{J}$ than the charge channel. Figure \ref{fig:K_matrix_elements} (a), describing the $U=0.1$ line in Figure \ref{fig:phase_diag} (a), demonstrates these features, as augmenting $K_{11}$, diminishing $K_{11}^{-1}$, and increasing $|K_{11}^{-1}|$ cause the TS states to decay slower than the SDW state, even for $U>0$.

Figure \ref{fig:phase_diag} (c) outlines the phase diagram of the EHB model with $V=0.1$. Comparing with the system in Figure \ref{fig:phase_diag} (a), the additional repulsive interaction shifts the phase diagram to the left. For smaller values of $|U|$, $V$ also induces two distinct CDWs, with wave numbers $2k_{\mathrm{F}}^\uparrow$ and $2k_{\mathrm{F}}^\downarrow$, each wave comprised of spin up or down electrons, respectively. The wave comprised of electrons with spin aligned opposite to the effective magnetic field is preferred for small $\bar{J}$, while the parallel case is favored for larger $\bar{J}$. This transition occurs as one of the spin bands approaches half-filling. 
Since Umklapp scattering is not accounted for in our model, further work is needed to understand the CDW transition. Furthermore, the quantity deciding which CDW is preferred, $K_{12}$, is plotted in Figure \ref{fig:K_matrix_elements} (b), and exhibits small oscillations for $\bar{J}<0.5$, revealing that there is no large distinction between the two CDW decay rates.
We also note that $V>0$ introduces a region without any divergent response functions for intermediate values of $\bar{J}$, since most diagonal entries in both $K$ and $K^{-1}$ are larger than one in Figure \ref{fig:K_matrix_elements} (b). Lastly, we again observe that the spin-polarized TS is dominant close to the unstable region, well inside the repulsive region of the phase diagram.

\textit{Outlook.} Our results indicate that spin-polarized triplet correlations in a metal chain coupled to a spin chain persist despite including repulsive interactions between electrons. 
This suggests that fluctuations in the spin chain provide a mechanism for superconductivity, as has been observed in similar planar interfaces \cite{Hugdal2018, Erlandsen2020}.
Our findings correspond well with the results found when coupling a metal chain to acoustical phonons. Strong electron-phonon coupling may induce superconductivity \cite{Loss1994}, particularly near the unstable region of the phase diagram. The spin fluctuations, however, change the spin structure of the electronic pairing compared to the phonon case, since they couple to the spin channel of the electrons.
In similar systems where helical edge states in topological insulators are coupled to spin impurities, it has been found that the coupling may cause Anderson localization of the edge states, suppressing transport \cite{Altshuler2013, Hsu2017}.
However, the backscattering that drives such systems into these insulating phases is absent in the system under consideration here due to the emergent effective magnetic field, leaving only the conventional insulating phases.

The system presented here is modeled using a TL description. There are several effects one could consider in future work, which would require an RG treatment. Among them are systematically accounting for Umklapp scattering, using a $SU(2)$ symmetric model, or removing the effective magnetic field by placing the metal chain between two spin chains. However, the emergent physics in our relatively simple, one-dimensional system still offers obvious parallels to magnon-mediated superconductivity in heterostructures of higher dimensions. Our main point is that we have demonstrated that spin-electron coupling provides a mechanism for driving superconducting instabilities even in non-Fermi liquids.

\textit{Acknowledgements.} This work was supported by the Research Council of Norway (RCN) through its Centres of Excellence funding scheme, Project No. 262633, "QuSpin", as well as RCN Project No. 323766.

\bibliography{main.bib}

\begin{thebibliography}{78}%
\makeatletter
\providecommand \@ifxundefined [1]{%
 \@ifx{#1\undefined}
}%
\providecommand \@ifnum [1]{%
 \ifnum #1\expandafter \@firstoftwo
 \else \expandafter \@secondoftwo
 \fi
}%
\providecommand \@ifx [1]{%
 \ifx #1\expandafter \@firstoftwo
 \else \expandafter \@secondoftwo
 \fi
}%
\providecommand \natexlab [1]{#1}%
\providecommand \enquote  [1]{``#1''}%
\providecommand \bibnamefont  [1]{#1}%
\providecommand \bibfnamefont [1]{#1}%
\providecommand \citenamefont [1]{#1}%
\providecommand \href@noop [0]{\@secondoftwo}%
\providecommand \href [0]{\begingroup \@sanitize@url \@href}%
\providecommand \@href[1]{\@@startlink{#1}\@@href}%
\providecommand \@@href[1]{\endgroup#1\@@endlink}%
\providecommand \@sanitize@url [0]{\catcode `\\12\catcode `\$12\catcode
  `\&12\catcode `\#12\catcode `\^12\catcode `\_12\catcode `\%12\relax}%
\providecommand \@@startlink[1]{}%
\providecommand \@@endlink[0]{}%
\providecommand \url  [0]{\begingroup\@sanitize@url \@url }%
\providecommand \@url [1]{\endgroup\@href {#1}{\urlprefix }}%
\providecommand \urlprefix  [0]{URL }%
\providecommand \Eprint [0]{\href }%
\providecommand \doibase [0]{https://doi.org/}%
\providecommand \selectlanguage [0]{\@gobble}%
\providecommand \bibinfo  [0]{\@secondoftwo}%
\providecommand \bibfield  [0]{\@secondoftwo}%
\providecommand \translation [1]{[#1]}%
\providecommand \BibitemOpen [0]{}%
\providecommand \bibitemStop [0]{}%
\providecommand \bibitemNoStop [0]{.\EOS\space}%
\providecommand \EOS [0]{\spacefactor3000\relax}%
\providecommand \BibitemShut  [1]{\csname bibitem#1\endcsname}%
\let\auto@bib@innerbib\@empty
\bibitem [{\citenamefont {Geim}\ and\ \citenamefont
  {Grigorieva}(2013)}]{Geim2013}%
  \BibitemOpen
  \bibfield  {author} {\bibinfo {author} {\bibfnamefont {A.~K.}\ \bibnamefont
  {Geim}}\ and\ \bibinfo {author} {\bibfnamefont {I.~V.}\ \bibnamefont
  {Grigorieva}},\ }\href {https://doi.org/10.1038/nature12385} {\bibfield
  {journal} {\bibinfo  {journal} {Nature}\ }\textbf {\bibinfo {volume} {499}},\
  \bibinfo {pages} {419} (\bibinfo {year} {2013})}\BibitemShut {NoStop}%
\bibitem [{\citenamefont {Aspelmeyer}\ \emph {et~al.}(2014)\citenamefont
  {Aspelmeyer}, \citenamefont {Kippenberg},\ and\ \citenamefont
  {Marquardt}}]{Aspelmeyer2014}%
  \BibitemOpen
  \bibfield  {author} {\bibinfo {author} {\bibfnamefont {M.}~\bibnamefont
  {Aspelmeyer}}, \bibinfo {author} {\bibfnamefont {T.~J.}\ \bibnamefont
  {Kippenberg}},\ and\ \bibinfo {author} {\bibfnamefont {F.}~\bibnamefont
  {Marquardt}},\ }\href {https://doi.org/10.1103/RevModPhys.86.1391} {\bibfield
   {journal} {\bibinfo  {journal} {Rev. Mod. Phys.}\ }\textbf {\bibinfo
  {volume} {86}},\ \bibinfo {pages} {1391} (\bibinfo {year}
  {2014})}\BibitemShut {NoStop}%
\bibitem [{\citenamefont {Clerk}\ \emph {et~al.}(2020)\citenamefont {Clerk},
  \citenamefont {Lehnert}, \citenamefont {Bertet}, \citenamefont {Petta},\ and\
  \citenamefont {Nakamura}}]{Clerk2020}%
  \BibitemOpen
  \bibfield  {author} {\bibinfo {author} {\bibfnamefont {A.~A.}\ \bibnamefont
  {Clerk}}, \bibinfo {author} {\bibfnamefont {K.~W.}\ \bibnamefont {Lehnert}},
  \bibinfo {author} {\bibfnamefont {P.}~\bibnamefont {Bertet}}, \bibinfo
  {author} {\bibfnamefont {J.~R.}\ \bibnamefont {Petta}},\ and\ \bibinfo
  {author} {\bibfnamefont {Y.}~\bibnamefont {Nakamura}},\ }\href
  {https://doi.org/10.1038/s41567-020-0797-9} {\bibfield  {journal} {\bibinfo
  {journal} {Nature Physics}\ }\textbf {\bibinfo {volume} {16}},\ \bibinfo
  {pages} {257} (\bibinfo {year} {2020})}\BibitemShut {NoStop}%
\bibitem [{\citenamefont {de~la Torre}\ \emph {et~al.}(2021)\citenamefont
  {de~la Torre}, \citenamefont {Kennes}, \citenamefont {Claassen},
  \citenamefont {Gerber}, \citenamefont {McIver},\ and\ \citenamefont
  {Sentef}}]{Torre2021}%
  \BibitemOpen
  \bibfield  {author} {\bibinfo {author} {\bibfnamefont {A.}~\bibnamefont
  {de~la Torre}}, \bibinfo {author} {\bibfnamefont {D.~M.}\ \bibnamefont
  {Kennes}}, \bibinfo {author} {\bibfnamefont {M.}~\bibnamefont {Claassen}},
  \bibinfo {author} {\bibfnamefont {S.}~\bibnamefont {Gerber}}, \bibinfo
  {author} {\bibfnamefont {J.~W.}\ \bibnamefont {McIver}},\ and\ \bibinfo
  {author} {\bibfnamefont {M.~A.}\ \bibnamefont {Sentef}},\ }\href
  {https://doi.org/10.1103/RevModPhys.93.041002} {\bibfield  {journal}
  {\bibinfo  {journal} {Rev. Mod. Phys.}\ }\textbf {\bibinfo {volume} {93}},\
  \bibinfo {pages} {041002} (\bibinfo {year} {2021})}\BibitemShut {NoStop}%
\bibitem [{\citenamefont {Bockrath}\ \emph {et~al.}(1999)\citenamefont
  {Bockrath}, \citenamefont {Cobden}, \citenamefont {Lu}, \citenamefont
  {Rinzler}, \citenamefont {Smalley}, \citenamefont {Balents},\ and\
  \citenamefont {McEuen}}]{Bockrath1999}%
  \BibitemOpen
  \bibfield  {author} {\bibinfo {author} {\bibfnamefont {M.}~\bibnamefont
  {Bockrath}}, \bibinfo {author} {\bibfnamefont {D.~H.}\ \bibnamefont
  {Cobden}}, \bibinfo {author} {\bibfnamefont {J.}~\bibnamefont {Lu}}, \bibinfo
  {author} {\bibfnamefont {A.~G.}\ \bibnamefont {Rinzler}}, \bibinfo {author}
  {\bibfnamefont {R.~E.}\ \bibnamefont {Smalley}}, \bibinfo {author}
  {\bibfnamefont {L.}~\bibnamefont {Balents}},\ and\ \bibinfo {author}
  {\bibfnamefont {P.~L.}\ \bibnamefont {McEuen}},\ }\href
  {https://doi.org/10.1038/17569} {\bibfield  {journal} {\bibinfo  {journal}
  {Nature}\ }\textbf {\bibinfo {volume} {397}},\ \bibinfo {pages} {598}
  (\bibinfo {year} {1999})}\BibitemShut {NoStop}%
\bibitem [{\citenamefont {Jotzu}\ \emph {et~al.}(2014)\citenamefont {Jotzu},
  \citenamefont {Messer}, \citenamefont {Desbuquois}, \citenamefont {Lebrat},
  \citenamefont {Uehlinger}, \citenamefont {Greif},\ and\ \citenamefont
  {Esslinger}}]{Jotzu2014}%
  \BibitemOpen
  \bibfield  {author} {\bibinfo {author} {\bibfnamefont {G.}~\bibnamefont
  {Jotzu}}, \bibinfo {author} {\bibfnamefont {M.}~\bibnamefont {Messer}},
  \bibinfo {author} {\bibfnamefont {R.}~\bibnamefont {Desbuquois}}, \bibinfo
  {author} {\bibfnamefont {M.}~\bibnamefont {Lebrat}}, \bibinfo {author}
  {\bibfnamefont {T.}~\bibnamefont {Uehlinger}}, \bibinfo {author}
  {\bibfnamefont {D.}~\bibnamefont {Greif}},\ and\ \bibinfo {author}
  {\bibfnamefont {T.}~\bibnamefont {Esslinger}},\ }\href
  {https://doi.org/10.1038/nature13915} {\bibfield  {journal} {\bibinfo
  {journal} {Nature}\ }\textbf {\bibinfo {volume} {515}},\ \bibinfo {pages}
  {237} (\bibinfo {year} {2014})}\BibitemShut {NoStop}%
\bibitem [{\citenamefont {Toskovic}\ \emph {et~al.}(2016)\citenamefont
  {Toskovic}, \citenamefont {van~den Berg}, \citenamefont {Spinelli},
  \citenamefont {Eliens}, \citenamefont {van~den Toorn}, \citenamefont
  {Bryant}, \citenamefont {Caux},\ and\ \citenamefont {Otte}}]{Toskovic2016}%
  \BibitemOpen
  \bibfield  {author} {\bibinfo {author} {\bibfnamefont {R.}~\bibnamefont
  {Toskovic}}, \bibinfo {author} {\bibfnamefont {R.}~\bibnamefont {van~den
  Berg}}, \bibinfo {author} {\bibfnamefont {A.}~\bibnamefont {Spinelli}},
  \bibinfo {author} {\bibfnamefont {I.~S.}\ \bibnamefont {Eliens}}, \bibinfo
  {author} {\bibfnamefont {B.}~\bibnamefont {van~den Toorn}}, \bibinfo {author}
  {\bibfnamefont {B.}~\bibnamefont {Bryant}}, \bibinfo {author} {\bibfnamefont
  {J.-S.}\ \bibnamefont {Caux}},\ and\ \bibinfo {author} {\bibfnamefont
  {A.~F.}\ \bibnamefont {Otte}},\ }\href {https://doi.org/10.1038/nphys3722}
  {\bibfield  {journal} {\bibinfo  {journal} {Nature Physics}\ }\textbf
  {\bibinfo {volume} {12}},\ \bibinfo {pages} {656} (\bibinfo {year}
  {2016})}\BibitemShut {NoStop}%
\bibitem [{\citenamefont {Sau}\ \emph {et~al.}(2010)\citenamefont {Sau},
  \citenamefont {Tewari}, \citenamefont {Lutchyn}, \citenamefont {Stanescu},\
  and\ \citenamefont {Das~Sarma}}]{Sau2010}%
  \BibitemOpen
  \bibfield  {author} {\bibinfo {author} {\bibfnamefont {J.~D.}\ \bibnamefont
  {Sau}}, \bibinfo {author} {\bibfnamefont {S.}~\bibnamefont {Tewari}},
  \bibinfo {author} {\bibfnamefont {R.~M.}\ \bibnamefont {Lutchyn}}, \bibinfo
  {author} {\bibfnamefont {T.~D.}\ \bibnamefont {Stanescu}},\ and\ \bibinfo
  {author} {\bibfnamefont {S.}~\bibnamefont {Das~Sarma}},\ }\href
  {https://doi.org/10.1103/PhysRevB.82.214509} {\bibfield  {journal} {\bibinfo
  {journal} {Phys. Rev. B}\ }\textbf {\bibinfo {volume} {82}},\ \bibinfo
  {pages} {214509} (\bibinfo {year} {2010})}\BibitemShut {NoStop}%
\bibitem [{\citenamefont {Amundsen}\ \emph {et~al.}(2018)\citenamefont
  {Amundsen}, \citenamefont {Hugdal}, \citenamefont {Sudb\o{}},\ and\
  \citenamefont {Linder}}]{Amundsen2018}%
  \BibitemOpen
  \bibfield  {author} {\bibinfo {author} {\bibfnamefont {M.}~\bibnamefont
  {Amundsen}}, \bibinfo {author} {\bibfnamefont {H.~G.}\ \bibnamefont
  {Hugdal}}, \bibinfo {author} {\bibfnamefont {A.}~\bibnamefont {Sudb\o{}}},\
  and\ \bibinfo {author} {\bibfnamefont {J.}~\bibnamefont {Linder}},\ }\href
  {https://doi.org/10.1103/PhysRevB.98.144505} {\bibfield  {journal} {\bibinfo
  {journal} {Phys. Rev. B}\ }\textbf {\bibinfo {volume} {98}},\ \bibinfo
  {pages} {144505} (\bibinfo {year} {2018})}\BibitemShut {NoStop}%
\bibitem [{\citenamefont {Rex}\ \emph {et~al.}(2020)\citenamefont {Rex},
  \citenamefont {Gornyi},\ and\ \citenamefont {Mirlin}}]{Rex2020}%
  \BibitemOpen
  \bibfield  {author} {\bibinfo {author} {\bibfnamefont {S.}~\bibnamefont
  {Rex}}, \bibinfo {author} {\bibfnamefont {I.~V.}\ \bibnamefont {Gornyi}},\
  and\ \bibinfo {author} {\bibfnamefont {A.~D.}\ \bibnamefont {Mirlin}},\
  }\href {https://doi.org/10.1103/PhysRevB.102.224501} {\bibfield  {journal}
  {\bibinfo  {journal} {Phys. Rev. B}\ }\textbf {\bibinfo {volume} {102}},\
  \bibinfo {pages} {224501} (\bibinfo {year} {2020})}\BibitemShut {NoStop}%
\bibitem [{\citenamefont {D\'{\i}az}\ \emph {et~al.}(2021)\citenamefont
  {D\'{\i}az}, \citenamefont {Klinovaja}, \citenamefont {Loss},\ and\
  \citenamefont {Hoffman}}]{Diaz2021}%
  \BibitemOpen
  \bibfield  {author} {\bibinfo {author} {\bibfnamefont {S.~A.}\ \bibnamefont
  {D\'{\i}az}}, \bibinfo {author} {\bibfnamefont {J.}~\bibnamefont
  {Klinovaja}}, \bibinfo {author} {\bibfnamefont {D.}~\bibnamefont {Loss}},\
  and\ \bibinfo {author} {\bibfnamefont {S.}~\bibnamefont {Hoffman}},\ }\href
  {https://doi.org/10.1103/PhysRevB.104.214501} {\bibfield  {journal} {\bibinfo
   {journal} {Phys. Rev. B}\ }\textbf {\bibinfo {volume} {104}},\ \bibinfo
  {pages} {214501} (\bibinfo {year} {2021})}\BibitemShut {NoStop}%
\bibitem [{\citenamefont {Gibertini}\ \emph {et~al.}(2019)\citenamefont
  {Gibertini}, \citenamefont {Koperski}, \citenamefont {Morpurgo},\ and\
  \citenamefont {Novoselov}}]{Gibertini2019}%
  \BibitemOpen
  \bibfield  {author} {\bibinfo {author} {\bibfnamefont {M.}~\bibnamefont
  {Gibertini}}, \bibinfo {author} {\bibfnamefont {M.}~\bibnamefont {Koperski}},
  \bibinfo {author} {\bibfnamefont {A.~F.}\ \bibnamefont {Morpurgo}},\ and\
  \bibinfo {author} {\bibfnamefont {K.~S.}\ \bibnamefont {Novoselov}},\ }\href
  {https://doi.org/10.1038/s41565-019-0438-6} {\bibfield  {journal} {\bibinfo
  {journal} {Nature Nanotechnology}\ }\textbf {\bibinfo {volume} {14}},\
  \bibinfo {pages} {408} (\bibinfo {year} {2019})}\BibitemShut {NoStop}%
\bibitem [{\citenamefont {Liu}\ \emph {et~al.}(2020)\citenamefont {Liu},
  \citenamefont {Kally}, \citenamefont {Pillsbury}, \citenamefont {Liu},
  \citenamefont {Chang}, \citenamefont {Ding}, \citenamefont {Cheng},
  \citenamefont {Hilse}, \citenamefont {Engel-Herbert}, \citenamefont
  {Richardella}, \citenamefont {Samarth},\ and\ \citenamefont {Wu}}]{Liu2020}%
  \BibitemOpen
  \bibfield  {author} {\bibinfo {author} {\bibfnamefont {T.}~\bibnamefont
  {Liu}}, \bibinfo {author} {\bibfnamefont {J.}~\bibnamefont {Kally}}, \bibinfo
  {author} {\bibfnamefont {T.}~\bibnamefont {Pillsbury}}, \bibinfo {author}
  {\bibfnamefont {C.}~\bibnamefont {Liu}}, \bibinfo {author} {\bibfnamefont
  {H.}~\bibnamefont {Chang}}, \bibinfo {author} {\bibfnamefont
  {J.}~\bibnamefont {Ding}}, \bibinfo {author} {\bibfnamefont {Y.}~\bibnamefont
  {Cheng}}, \bibinfo {author} {\bibfnamefont {M.}~\bibnamefont {Hilse}},
  \bibinfo {author} {\bibfnamefont {R.}~\bibnamefont {Engel-Herbert}}, \bibinfo
  {author} {\bibfnamefont {A.}~\bibnamefont {Richardella}}, \bibinfo {author}
  {\bibfnamefont {N.}~\bibnamefont {Samarth}},\ and\ \bibinfo {author}
  {\bibfnamefont {M.}~\bibnamefont {Wu}},\ }\href
  {https://doi.org/10.1103/PhysRevLett.125.017204} {\bibfield  {journal}
  {\bibinfo  {journal} {Phys. Rev. Lett.}\ }\textbf {\bibinfo {volume} {125}},\
  \bibinfo {pages} {017204} (\bibinfo {year} {2020})}\BibitemShut {NoStop}%
\bibitem [{\citenamefont {Deng}\ \emph {et~al.}(2020)\citenamefont {Deng},
  \citenamefont {Yu}, \citenamefont {Shi}, \citenamefont {Guo}, \citenamefont
  {Xu}, \citenamefont {Wang}, \citenamefont {Chen},\ and\ \citenamefont
  {Zhang}}]{Deng2020}%
  \BibitemOpen
  \bibfield  {author} {\bibinfo {author} {\bibfnamefont {Y.}~\bibnamefont
  {Deng}}, \bibinfo {author} {\bibfnamefont {Y.}~\bibnamefont {Yu}}, \bibinfo
  {author} {\bibfnamefont {M.~Z.}\ \bibnamefont {Shi}}, \bibinfo {author}
  {\bibfnamefont {Z.}~\bibnamefont {Guo}}, \bibinfo {author} {\bibfnamefont
  {Z.}~\bibnamefont {Xu}}, \bibinfo {author} {\bibfnamefont {J.}~\bibnamefont
  {Wang}}, \bibinfo {author} {\bibfnamefont {X.~H.}\ \bibnamefont {Chen}},\
  and\ \bibinfo {author} {\bibfnamefont {Y.}~\bibnamefont {Zhang}},\ }\href
  {https://doi.org/10.1126/science.aax8156} {\bibfield  {journal} {\bibinfo
  {journal} {Science}\ }\textbf {\bibinfo {volume} {367}},\ \bibinfo {pages}
  {895} (\bibinfo {year} {2020})}\BibitemShut {NoStop}%
\bibitem [{\citenamefont {König}\ \emph {et~al.}(2007)\citenamefont {König},
  \citenamefont {Wiedmann}, \citenamefont {Brüne}, \citenamefont {Roth},
  \citenamefont {Buhmann}, \citenamefont {Molenkamp}, \citenamefont {Qi},\ and\
  \citenamefont {Zhang}}]{Konig2007}%
  \BibitemOpen
  \bibfield  {author} {\bibinfo {author} {\bibfnamefont {M.}~\bibnamefont
  {König}}, \bibinfo {author} {\bibfnamefont {S.}~\bibnamefont {Wiedmann}},
  \bibinfo {author} {\bibfnamefont {C.}~\bibnamefont {Brüne}}, \bibinfo
  {author} {\bibfnamefont {A.}~\bibnamefont {Roth}}, \bibinfo {author}
  {\bibfnamefont {H.}~\bibnamefont {Buhmann}}, \bibinfo {author} {\bibfnamefont
  {L.~W.}\ \bibnamefont {Molenkamp}}, \bibinfo {author} {\bibfnamefont {X.-L.}\
  \bibnamefont {Qi}},\ and\ \bibinfo {author} {\bibfnamefont {S.-C.}\
  \bibnamefont {Zhang}},\ }\href {https://doi.org/10.1126/science.1148047}
  {\bibfield  {journal} {\bibinfo  {journal} {Science}\ }\textbf {\bibinfo
  {volume} {318}},\ \bibinfo {pages} {766} (\bibinfo {year}
  {2007})}\BibitemShut {NoStop}%
\bibitem [{\citenamefont {Hsieh}\ \emph {et~al.}(2008)\citenamefont {Hsieh},
  \citenamefont {Qian}, \citenamefont {Wray}, \citenamefont {Xia},
  \citenamefont {Hor}, \citenamefont {Cava},\ and\ \citenamefont
  {Hasan}}]{Hsieh2008}%
  \BibitemOpen
  \bibfield  {author} {\bibinfo {author} {\bibfnamefont {D.}~\bibnamefont
  {Hsieh}}, \bibinfo {author} {\bibfnamefont {D.}~\bibnamefont {Qian}},
  \bibinfo {author} {\bibfnamefont {L.}~\bibnamefont {Wray}}, \bibinfo {author}
  {\bibfnamefont {Y.}~\bibnamefont {Xia}}, \bibinfo {author} {\bibfnamefont
  {Y.~S.}\ \bibnamefont {Hor}}, \bibinfo {author} {\bibfnamefont {R.~J.}\
  \bibnamefont {Cava}},\ and\ \bibinfo {author} {\bibfnamefont {M.~Z.}\
  \bibnamefont {Hasan}},\ }\href {https://doi.org/10.1038/nature06843}
  {\bibfield  {journal} {\bibinfo  {journal} {Nature}\ }\textbf {\bibinfo
  {volume} {452}},\ \bibinfo {pages} {970} (\bibinfo {year}
  {2008})}\BibitemShut {NoStop}%
\bibitem [{\citenamefont {Hasan}\ and\ \citenamefont {Kane}(2010)}]{Hasan2010}%
  \BibitemOpen
  \bibfield  {author} {\bibinfo {author} {\bibfnamefont {M.~Z.}\ \bibnamefont
  {Hasan}}\ and\ \bibinfo {author} {\bibfnamefont {C.~L.}\ \bibnamefont
  {Kane}},\ }\href {https://doi.org/10.1103/RevModPhys.82.3045} {\bibfield
  {journal} {\bibinfo  {journal} {Rev. Mod. Phys.}\ }\textbf {\bibinfo {volume}
  {82}},\ \bibinfo {pages} {3045} (\bibinfo {year} {2010})}\BibitemShut
  {NoStop}%
\bibitem [{\citenamefont {Kitaev}(2001)}]{Kitaev2001}%
  \BibitemOpen
  \bibfield  {author} {\bibinfo {author} {\bibfnamefont {A.~Y.}\ \bibnamefont
  {Kitaev}},\ }\href {https://doi.org/10.1070/1063-7869/44/10s/s29} {\bibfield
  {journal} {\bibinfo  {journal} {Physics-Uspekhi}\ }\textbf {\bibinfo {volume}
  {44}},\ \bibinfo {pages} {131} (\bibinfo {year} {2001})}\BibitemShut
  {NoStop}%
\bibitem [{\citenamefont {Mourik}\ \emph {et~al.}(2012)\citenamefont {Mourik},
  \citenamefont {Zuo}, \citenamefont {Frolov}, \citenamefont {Plissard},
  \citenamefont {Bakkers},\ and\ \citenamefont {Kouwenhoven}}]{Mourik2012}%
  \BibitemOpen
  \bibfield  {author} {\bibinfo {author} {\bibfnamefont {V.}~\bibnamefont
  {Mourik}}, \bibinfo {author} {\bibfnamefont {K.}~\bibnamefont {Zuo}},
  \bibinfo {author} {\bibfnamefont {S.~M.}\ \bibnamefont {Frolov}}, \bibinfo
  {author} {\bibfnamefont {S.~R.}\ \bibnamefont {Plissard}}, \bibinfo {author}
  {\bibfnamefont {E.~P. A.~M.}\ \bibnamefont {Bakkers}},\ and\ \bibinfo
  {author} {\bibfnamefont {L.~P.}\ \bibnamefont {Kouwenhoven}},\ }\href
  {https://doi.org/10.1126/science.1222360} {\bibfield  {journal} {\bibinfo
  {journal} {Science}\ }\textbf {\bibinfo {volume} {336}},\ \bibinfo {pages}
  {1003} (\bibinfo {year} {2012})}\BibitemShut {NoStop}%
\bibitem [{\citenamefont {Albrecht}\ \emph {et~al.}(2016)\citenamefont
  {Albrecht}, \citenamefont {Higginbotham}, \citenamefont {Madsen},
  \citenamefont {Kuemmeth}, \citenamefont {Jespersen}, \citenamefont
  {Nyg{\aa}rd}, \citenamefont {Krogstrup},\ and\ \citenamefont
  {Marcus}}]{Albrecht2016}%
  \BibitemOpen
  \bibfield  {author} {\bibinfo {author} {\bibfnamefont {S.~M.}\ \bibnamefont
  {Albrecht}}, \bibinfo {author} {\bibfnamefont {A.~P.}\ \bibnamefont
  {Higginbotham}}, \bibinfo {author} {\bibfnamefont {M.}~\bibnamefont
  {Madsen}}, \bibinfo {author} {\bibfnamefont {F.}~\bibnamefont {Kuemmeth}},
  \bibinfo {author} {\bibfnamefont {T.~S.}\ \bibnamefont {Jespersen}}, \bibinfo
  {author} {\bibfnamefont {J.}~\bibnamefont {Nyg{\aa}rd}}, \bibinfo {author}
  {\bibfnamefont {P.}~\bibnamefont {Krogstrup}},\ and\ \bibinfo {author}
  {\bibfnamefont {C.~M.}\ \bibnamefont {Marcus}},\ }\href
  {https://doi.org/10.1038/nature17162} {\bibfield  {journal} {\bibinfo
  {journal} {Nature}\ }\textbf {\bibinfo {volume} {531}},\ \bibinfo {pages}
  {206} (\bibinfo {year} {2016})}\BibitemShut {NoStop}%
\bibitem [{\citenamefont {Liu}\ \emph {et~al.}(2017)\citenamefont {Liu},
  \citenamefont {Sau}, \citenamefont {Stanescu},\ and\ \citenamefont
  {Das~Sarma}}]{Liu2017}%
  \BibitemOpen
  \bibfield  {author} {\bibinfo {author} {\bibfnamefont {C.-X.}\ \bibnamefont
  {Liu}}, \bibinfo {author} {\bibfnamefont {J.~D.}\ \bibnamefont {Sau}},
  \bibinfo {author} {\bibfnamefont {T.~D.}\ \bibnamefont {Stanescu}},\ and\
  \bibinfo {author} {\bibfnamefont {S.}~\bibnamefont {Das~Sarma}},\ }\href
  {https://doi.org/10.1103/PhysRevB.96.075161} {\bibfield  {journal} {\bibinfo
  {journal} {Phys. Rev. B}\ }\textbf {\bibinfo {volume} {96}},\ \bibinfo
  {pages} {075161} (\bibinfo {year} {2017})}\BibitemShut {NoStop}%
\bibitem [{\citenamefont {Lutchyn}\ \emph {et~al.}(2018)\citenamefont
  {Lutchyn}, \citenamefont {Bakkers}, \citenamefont {Kouwenhoven},
  \citenamefont {Krogstrup}, \citenamefont {Marcus},\ and\ \citenamefont
  {Oreg}}]{Lutchyn2018}%
  \BibitemOpen
  \bibfield  {author} {\bibinfo {author} {\bibfnamefont {R.~M.}\ \bibnamefont
  {Lutchyn}}, \bibinfo {author} {\bibfnamefont {E.~P. A.~M.}\ \bibnamefont
  {Bakkers}}, \bibinfo {author} {\bibfnamefont {L.~P.}\ \bibnamefont
  {Kouwenhoven}}, \bibinfo {author} {\bibfnamefont {P.}~\bibnamefont
  {Krogstrup}}, \bibinfo {author} {\bibfnamefont {C.~M.}\ \bibnamefont
  {Marcus}},\ and\ \bibinfo {author} {\bibfnamefont {Y.}~\bibnamefont {Oreg}},\
  }\href {https://doi.org/10.1038/s41578-018-0003-1} {\bibfield  {journal}
  {\bibinfo  {journal} {Nature Reviews Materials}\ }\textbf {\bibinfo {volume}
  {3}},\ \bibinfo {pages} {52} (\bibinfo {year} {2018})}\BibitemShut {NoStop}%
\bibitem [{\citenamefont {Zhang}\ \emph {et~al.}(2019)\citenamefont {Zhang},
  \citenamefont {Liu}, \citenamefont {Wimmer},\ and\ \citenamefont
  {Kouwenhoven}}]{Zhang2019}%
  \BibitemOpen
  \bibfield  {author} {\bibinfo {author} {\bibfnamefont {H.}~\bibnamefont
  {Zhang}}, \bibinfo {author} {\bibfnamefont {D.~E.}\ \bibnamefont {Liu}},
  \bibinfo {author} {\bibfnamefont {M.}~\bibnamefont {Wimmer}},\ and\ \bibinfo
  {author} {\bibfnamefont {L.~P.}\ \bibnamefont {Kouwenhoven}},\ }\href
  {https://doi.org/10.1038/s41467-019-13133-1} {\bibfield  {journal} {\bibinfo
  {journal} {Nature Communications}\ }\textbf {\bibinfo {volume} {10}},\
  \bibinfo {pages} {5128} (\bibinfo {year} {2019})}\BibitemShut {NoStop}%
\bibitem [{\citenamefont {Vaitiek{\.{e}}nas}\ \emph {et~al.}(2021)\citenamefont
  {Vaitiek{\.{e}}nas}, \citenamefont {Liu}, \citenamefont {Krogstrup},\ and\
  \citenamefont {Marcus}}]{Vaitiekenas2021}%
  \BibitemOpen
  \bibfield  {author} {\bibinfo {author} {\bibfnamefont {S.}~\bibnamefont
  {Vaitiek{\.{e}}nas}}, \bibinfo {author} {\bibfnamefont {Y.}~\bibnamefont
  {Liu}}, \bibinfo {author} {\bibfnamefont {P.}~\bibnamefont {Krogstrup}},\
  and\ \bibinfo {author} {\bibfnamefont {C.~M.}\ \bibnamefont {Marcus}},\
  }\href {https://doi.org/10.1038/s41567-020-1017-3} {\bibfield  {journal}
  {\bibinfo  {journal} {Nature Physics}\ }\textbf {\bibinfo {volume} {17}},\
  \bibinfo {pages} {43} (\bibinfo {year} {2021})}\BibitemShut {NoStop}%
\bibitem [{\citenamefont {Flensberg}\ \emph {et~al.}(2021)\citenamefont
  {Flensberg}, \citenamefont {von Oppen},\ and\ \citenamefont
  {Stern}}]{Flensberg2021}%
  \BibitemOpen
  \bibfield  {author} {\bibinfo {author} {\bibfnamefont {K.}~\bibnamefont
  {Flensberg}}, \bibinfo {author} {\bibfnamefont {F.}~\bibnamefont {von
  Oppen}},\ and\ \bibinfo {author} {\bibfnamefont {A.}~\bibnamefont {Stern}},\
  }\href {https://doi.org/10.1038/s41578-021-00336-6} {\bibfield  {journal}
  {\bibinfo  {journal} {Nature Reviews Materials}\ }\textbf {\bibinfo {volume}
  {6}},\ \bibinfo {pages} {944} (\bibinfo {year} {2021})}\BibitemShut {NoStop}%
\bibitem [{\citenamefont {Pan}\ and\ \citenamefont
  {Das~Sarma}(2022)}]{Pan2022}%
  \BibitemOpen
  \bibfield  {author} {\bibinfo {author} {\bibfnamefont {H.}~\bibnamefont
  {Pan}}\ and\ \bibinfo {author} {\bibfnamefont {S.}~\bibnamefont
  {Das~Sarma}},\ }\href {https://doi.org/10.1103/PhysRevB.105.115432}
  {\bibfield  {journal} {\bibinfo  {journal} {Phys. Rev. B}\ }\textbf {\bibinfo
  {volume} {105}},\ \bibinfo {pages} {115432} (\bibinfo {year}
  {2022})}\BibitemShut {NoStop}%
\bibitem [{\citenamefont {Escribano}\ \emph {et~al.}(2022)\citenamefont
  {Escribano}, \citenamefont {Maiani}, \citenamefont {Leijnse}, \citenamefont
  {Flensberg}, \citenamefont {Oreg}, \citenamefont {Levy~Yeyati}, \citenamefont
  {Prada},\ and\ \citenamefont {Seoane~Souto}}]{Escribano2022}%
  \BibitemOpen
  \bibfield  {author} {\bibinfo {author} {\bibfnamefont {S.~D.}\ \bibnamefont
  {Escribano}}, \bibinfo {author} {\bibfnamefont {A.}~\bibnamefont {Maiani}},
  \bibinfo {author} {\bibfnamefont {M.}~\bibnamefont {Leijnse}}, \bibinfo
  {author} {\bibfnamefont {K.}~\bibnamefont {Flensberg}}, \bibinfo {author}
  {\bibfnamefont {Y.}~\bibnamefont {Oreg}}, \bibinfo {author} {\bibfnamefont
  {A.}~\bibnamefont {Levy~Yeyati}}, \bibinfo {author} {\bibfnamefont
  {E.}~\bibnamefont {Prada}},\ and\ \bibinfo {author} {\bibfnamefont
  {R.}~\bibnamefont {Seoane~Souto}},\ }\href
  {https://doi.org/10.1038/s41535-022-00489-9} {\bibfield  {journal} {\bibinfo
  {journal} {npj Quantum Materials}\ }\textbf {\bibinfo {volume} {7}},\
  \bibinfo {pages} {81} (\bibinfo {year} {2022})}\BibitemShut {NoStop}%
\bibitem [{\citenamefont {Linder}\ and\ \citenamefont
  {Robinson}(2015)}]{Linder2015}%
  \BibitemOpen
  \bibfield  {author} {\bibinfo {author} {\bibfnamefont {J.}~\bibnamefont
  {Linder}}\ and\ \bibinfo {author} {\bibfnamefont {J.~W.~A.}\ \bibnamefont
  {Robinson}},\ }\href {https://doi.org/10.1038/nphys3242} {\bibfield
  {journal} {\bibinfo  {journal} {Nature Physics}\ }\textbf {\bibinfo {volume}
  {11}},\ \bibinfo {pages} {307} (\bibinfo {year} {2015})}\BibitemShut
  {NoStop}%
\bibitem [{\citenamefont {Eschrig}(2015)}]{Eschrig2015}%
  \BibitemOpen
  \bibfield  {author} {\bibinfo {author} {\bibfnamefont {M.}~\bibnamefont
  {Eschrig}},\ }\href {https://doi.org/10.1088/0034-4885/78/10/104501}
  {\bibfield  {journal} {\bibinfo  {journal} {Reports on Progress in Physics}\
  }\textbf {\bibinfo {volume} {78}},\ \bibinfo {pages} {104501} (\bibinfo
  {year} {2015})}\BibitemShut {NoStop}%
\bibitem [{\citenamefont {Kargarian}\ \emph {et~al.}(2016)\citenamefont
  {Kargarian}, \citenamefont {Efimkin},\ and\ \citenamefont
  {Galitski}}]{Kargarian2016}%
  \BibitemOpen
  \bibfield  {author} {\bibinfo {author} {\bibfnamefont {M.}~\bibnamefont
  {Kargarian}}, \bibinfo {author} {\bibfnamefont {D.~K.}\ \bibnamefont
  {Efimkin}},\ and\ \bibinfo {author} {\bibfnamefont {V.}~\bibnamefont
  {Galitski}},\ }\href {https://doi.org/10.1103/PhysRevLett.117.076806}
  {\bibfield  {journal} {\bibinfo  {journal} {Phys. Rev. Lett.}\ }\textbf
  {\bibinfo {volume} {117}},\ \bibinfo {pages} {076806} (\bibinfo {year}
  {2016})}\BibitemShut {NoStop}%
\bibitem [{\citenamefont {Hugdal}\ \emph {et~al.}(2018)\citenamefont {Hugdal},
  \citenamefont {Rex}, \citenamefont {Nogueira},\ and\ \citenamefont
  {Sudb\o{}}}]{Hugdal2018}%
  \BibitemOpen
  \bibfield  {author} {\bibinfo {author} {\bibfnamefont {H.~G.}\ \bibnamefont
  {Hugdal}}, \bibinfo {author} {\bibfnamefont {S.}~\bibnamefont {Rex}},
  \bibinfo {author} {\bibfnamefont {F.~S.}\ \bibnamefont {Nogueira}},\ and\
  \bibinfo {author} {\bibfnamefont {A.}~\bibnamefont {Sudb\o{}}},\ }\href
  {https://doi.org/10.1103/PhysRevB.97.195438} {\bibfield  {journal} {\bibinfo
  {journal} {Phys. Rev. B}\ }\textbf {\bibinfo {volume} {97}},\ \bibinfo
  {pages} {195438} (\bibinfo {year} {2018})}\BibitemShut {NoStop}%
\bibitem [{\citenamefont {Rohling}\ \emph {et~al.}(2018)\citenamefont
  {Rohling}, \citenamefont {Fj\ae{}rbu},\ and\ \citenamefont
  {Brataas}}]{Rohling2018}%
  \BibitemOpen
  \bibfield  {author} {\bibinfo {author} {\bibfnamefont {N.}~\bibnamefont
  {Rohling}}, \bibinfo {author} {\bibfnamefont {E.~L.}\ \bibnamefont
  {Fj\ae{}rbu}},\ and\ \bibinfo {author} {\bibfnamefont {A.}~\bibnamefont
  {Brataas}},\ }\href {https://doi.org/10.1103/PhysRevB.97.115401} {\bibfield
  {journal} {\bibinfo  {journal} {Phys. Rev. B}\ }\textbf {\bibinfo {volume}
  {97}},\ \bibinfo {pages} {115401} (\bibinfo {year} {2018})}\BibitemShut
  {NoStop}%
\bibitem [{\citenamefont {Fj\ae{}rbu}\ \emph {et~al.}(2019)\citenamefont
  {Fj\ae{}rbu}, \citenamefont {Rohling},\ and\ \citenamefont
  {Brataas}}]{Fjaerbu2019}%
  \BibitemOpen
  \bibfield  {author} {\bibinfo {author} {\bibfnamefont {E.~L.}\ \bibnamefont
  {Fj\ae{}rbu}}, \bibinfo {author} {\bibfnamefont {N.}~\bibnamefont
  {Rohling}},\ and\ \bibinfo {author} {\bibfnamefont {A.}~\bibnamefont
  {Brataas}},\ }\href {https://doi.org/10.1103/PhysRevB.100.125432} {\bibfield
  {journal} {\bibinfo  {journal} {Phys. Rev. B}\ }\textbf {\bibinfo {volume}
  {100}},\ \bibinfo {pages} {125432} (\bibinfo {year} {2019})}\BibitemShut
  {NoStop}%
\bibitem [{\citenamefont {Erlandsen}\ \emph {et~al.}(2019)\citenamefont
  {Erlandsen}, \citenamefont {Kamra}, \citenamefont {Brataas},\ and\
  \citenamefont {Sudb\o{}}}]{Erlandsen2019}%
  \BibitemOpen
  \bibfield  {author} {\bibinfo {author} {\bibfnamefont {E.}~\bibnamefont
  {Erlandsen}}, \bibinfo {author} {\bibfnamefont {A.}~\bibnamefont {Kamra}},
  \bibinfo {author} {\bibfnamefont {A.}~\bibnamefont {Brataas}},\ and\ \bibinfo
  {author} {\bibfnamefont {A.}~\bibnamefont {Sudb\o{}}},\ }\href
  {https://doi.org/10.1103/PhysRevB.100.100503} {\bibfield  {journal} {\bibinfo
   {journal} {Phys. Rev. B}\ }\textbf {\bibinfo {volume} {100}},\ \bibinfo
  {pages} {100503} (\bibinfo {year} {2019})}\BibitemShut {NoStop}%
\bibitem [{\citenamefont {Erlandsen}\ \emph {et~al.}(2020)\citenamefont
  {Erlandsen}, \citenamefont {Brataas},\ and\ \citenamefont
  {Sudb\o{}}}]{Erlandsen2020}%
  \BibitemOpen
  \bibfield  {author} {\bibinfo {author} {\bibfnamefont {E.}~\bibnamefont
  {Erlandsen}}, \bibinfo {author} {\bibfnamefont {A.}~\bibnamefont {Brataas}},\
  and\ \bibinfo {author} {\bibfnamefont {A.}~\bibnamefont {Sudb\o{}}},\ }\href
  {https://doi.org/10.1103/PhysRevB.101.094503} {\bibfield  {journal} {\bibinfo
   {journal} {Phys. Rev. B}\ }\textbf {\bibinfo {volume} {101}},\ \bibinfo
  {pages} {094503} (\bibinfo {year} {2020})}\BibitemShut {NoStop}%
\bibitem [{\citenamefont {Thingstad}\ \emph {et~al.}(2021)\citenamefont
  {Thingstad}, \citenamefont {Erlandsen},\ and\ \citenamefont
  {Sudb\o{}}}]{Thingstad2021}%
  \BibitemOpen
  \bibfield  {author} {\bibinfo {author} {\bibfnamefont {E.}~\bibnamefont
  {Thingstad}}, \bibinfo {author} {\bibfnamefont {E.}~\bibnamefont
  {Erlandsen}},\ and\ \bibinfo {author} {\bibfnamefont {A.}~\bibnamefont
  {Sudb\o{}}},\ }\href {https://doi.org/10.1103/PhysRevB.104.014508} {\bibfield
   {journal} {\bibinfo  {journal} {Phys. Rev. B}\ }\textbf {\bibinfo {volume}
  {104}},\ \bibinfo {pages} {014508} (\bibinfo {year} {2021})}\BibitemShut
  {NoStop}%
\bibitem [{\citenamefont {Tomonaga}(1950)}]{Tomonaga1950}%
  \BibitemOpen
  \bibfield  {author} {\bibinfo {author} {\bibfnamefont {S.-i.}\ \bibnamefont
  {Tomonaga}},\ }\href {https://doi.org/10.1143/ptp/5.4.544} {\bibfield
  {journal} {\bibinfo  {journal} {Progress of Theoretical Physics}\ }\textbf
  {\bibinfo {volume} {5}},\ \bibinfo {pages} {544} (\bibinfo {year}
  {1950})}\BibitemShut {NoStop}%
\bibitem [{\citenamefont {Luttinger}(1963)}]{Luttinger1963}%
  \BibitemOpen
  \bibfield  {author} {\bibinfo {author} {\bibfnamefont {J.~M.}\ \bibnamefont
  {Luttinger}},\ }\href {https://doi.org/10.1063/1.1704046} {\bibfield
  {journal} {\bibinfo  {journal} {Journal of Mathematical Physics}\ }\textbf
  {\bibinfo {volume} {4}},\ \bibinfo {pages} {1154} (\bibinfo {year}
  {1963})}\BibitemShut {NoStop}%
\bibitem [{\citenamefont {Haldane}(1981)}]{Haldane1981}%
  \BibitemOpen
  \bibfield  {author} {\bibinfo {author} {\bibfnamefont {F.~D.~M.}\
  \bibnamefont {Haldane}},\ }\href
  {https://doi.org/10.1088/0022-3719/14/19/010} {\bibfield  {journal} {\bibinfo
   {journal} {Journal of Physics C: Solid State Physics}\ }\textbf {\bibinfo
  {volume} {14}},\ \bibinfo {pages} {2585} (\bibinfo {year}
  {1981})}\BibitemShut {NoStop}%
\bibitem [{\citenamefont {Anderson}(1997)}]{Anderson1997}%
  \BibitemOpen
  \bibfield  {author} {\bibinfo {author} {\bibfnamefont {P.~W.}\ \bibnamefont
  {Anderson}},\ }\href@noop {} {\emph {\bibinfo {title} {The Theory of
  Superconductivity in the High-$T_c$ Cuprate Superconductors}}}\ (\bibinfo
  {publisher} {Princeton University Press},\ \bibinfo {year}
  {1997})\BibitemShut {NoStop}%
\bibitem [{\citenamefont {Lederer}\ \emph {et~al.}(2017)\citenamefont
  {Lederer}, \citenamefont {Schattner}, \citenamefont {Berg},\ and\
  \citenamefont {Kivelson}}]{Lederer2017}%
  \BibitemOpen
  \bibfield  {author} {\bibinfo {author} {\bibfnamefont {S.}~\bibnamefont
  {Lederer}}, \bibinfo {author} {\bibfnamefont {Y.}~\bibnamefont {Schattner}},
  \bibinfo {author} {\bibfnamefont {E.}~\bibnamefont {Berg}},\ and\ \bibinfo
  {author} {\bibfnamefont {S.~A.}\ \bibnamefont {Kivelson}},\ }\href
  {https://doi.org/10.1073/pnas.1620651114} {\bibfield  {journal} {\bibinfo
  {journal} {Proceedings of the National Academy of Sciences}\ }\textbf
  {\bibinfo {volume} {114}},\ \bibinfo {pages} {4905} (\bibinfo {year}
  {2017})}\BibitemShut {NoStop}%
\bibitem [{\citenamefont {Sandler}\ and\ \citenamefont
  {Maslov}(1997)}]{Sandler1997}%
  \BibitemOpen
  \bibfield  {author} {\bibinfo {author} {\bibfnamefont {N.~P.}\ \bibnamefont
  {Sandler}}\ and\ \bibinfo {author} {\bibfnamefont {D.~L.}\ \bibnamefont
  {Maslov}},\ }\href {https://doi.org/10.1103/PhysRevB.55.13808} {\bibfield
  {journal} {\bibinfo  {journal} {Phys. Rev. B}\ }\textbf {\bibinfo {volume}
  {55}},\ \bibinfo {pages} {13808} (\bibinfo {year} {1997})}\BibitemShut
  {NoStop}%
\bibitem [{\citenamefont {Mukhopadhyay}\ \emph {et~al.}(2001)\citenamefont
  {Mukhopadhyay}, \citenamefont {Kane},\ and\ \citenamefont
  {Lubensky}}]{Mukhopadhyay2001}%
  \BibitemOpen
  \bibfield  {author} {\bibinfo {author} {\bibfnamefont {R.}~\bibnamefont
  {Mukhopadhyay}}, \bibinfo {author} {\bibfnamefont {C.~L.}\ \bibnamefont
  {Kane}},\ and\ \bibinfo {author} {\bibfnamefont {T.~C.}\ \bibnamefont
  {Lubensky}},\ }\href {https://doi.org/10.1103/PhysRevB.64.045120} {\bibfield
  {journal} {\bibinfo  {journal} {Phys. Rev. B}\ }\textbf {\bibinfo {volume}
  {64}},\ \bibinfo {pages} {045120} (\bibinfo {year} {2001})}\BibitemShut
  {NoStop}%
\bibitem [{\citenamefont {Yurkevich}(2013)}]{Yurkevich2013}%
  \BibitemOpen
  \bibfield  {author} {\bibinfo {author} {\bibfnamefont {I.~V.}\ \bibnamefont
  {Yurkevich}},\ }\href {https://doi.org/10.1209/0295-5075/104/37004}
  {\bibfield  {journal} {\bibinfo  {journal} {{EPL} (Europhysics Letters)}\
  }\textbf {\bibinfo {volume} {104}},\ \bibinfo {pages} {37004} (\bibinfo
  {year} {2013})}\BibitemShut {NoStop}%
\bibitem [{\citenamefont {Kagalovsky}\ \emph {et~al.}(2017)\citenamefont
  {Kagalovsky}, \citenamefont {Lerner},\ and\ \citenamefont
  {Yurkevich}}]{Kagalovsky2017}%
  \BibitemOpen
  \bibfield  {author} {\bibinfo {author} {\bibfnamefont {V.}~\bibnamefont
  {Kagalovsky}}, \bibinfo {author} {\bibfnamefont {I.~V.}\ \bibnamefont
  {Lerner}},\ and\ \bibinfo {author} {\bibfnamefont {I.~V.}\ \bibnamefont
  {Yurkevich}},\ }\href {https://doi.org/10.1103/PhysRevB.95.205122} {\bibfield
   {journal} {\bibinfo  {journal} {Phys. Rev. B}\ }\textbf {\bibinfo {volume}
  {95}},\ \bibinfo {pages} {205122} (\bibinfo {year} {2017})}\BibitemShut
  {NoStop}%
\bibitem [{\citenamefont {Jones}\ \emph {et~al.}(2017)\citenamefont {Jones},
  \citenamefont {Lerner},\ and\ \citenamefont {Yurkevich}}]{Jones2017}%
  \BibitemOpen
  \bibfield  {author} {\bibinfo {author} {\bibfnamefont {M.}~\bibnamefont
  {Jones}}, \bibinfo {author} {\bibfnamefont {I.~V.}\ \bibnamefont {Lerner}},\
  and\ \bibinfo {author} {\bibfnamefont {I.~V.}\ \bibnamefont {Yurkevich}},\
  }\href {https://doi.org/10.1103/PhysRevB.96.174210} {\bibfield  {journal}
  {\bibinfo  {journal} {Phys. Rev. B}\ }\textbf {\bibinfo {volume} {96}},\
  \bibinfo {pages} {174210} (\bibinfo {year} {2017})}\BibitemShut {NoStop}%
\bibitem [{\citenamefont {Yurkevich}(2017)}]{Yurkevich2017}%
  \BibitemOpen
  \bibfield  {author} {\bibinfo {author} {\bibfnamefont {I.~V.}\ \bibnamefont
  {Yurkevich}},\ }\href {https://doi.org/10.1038/s41598-017-03823-5} {\bibfield
   {journal} {\bibinfo  {journal} {Scientific Reports}\ }\textbf {\bibinfo
  {volume} {7}},\ \bibinfo {pages} {3550} (\bibinfo {year} {2017})}\BibitemShut
  {NoStop}%
\bibitem [{\citenamefont {Voit}(1992)}]{Voit1992}%
  \BibitemOpen
  \bibfield  {author} {\bibinfo {author} {\bibfnamefont {J.}~\bibnamefont
  {Voit}},\ }\href {https://doi.org/10.1103/PhysRevB.45.4027} {\bibfield
  {journal} {\bibinfo  {journal} {Phys. Rev. B}\ }\textbf {\bibinfo {volume}
  {45}},\ \bibinfo {pages} {4027} (\bibinfo {year} {1992})}\BibitemShut
  {NoStop}%
\bibitem [{\citenamefont {Tsuchiizu}\ and\ \citenamefont
  {Furusaki}(2002)}]{Tsuchiizu2002}%
  \BibitemOpen
  \bibfield  {author} {\bibinfo {author} {\bibfnamefont {M.}~\bibnamefont
  {Tsuchiizu}}\ and\ \bibinfo {author} {\bibfnamefont {A.}~\bibnamefont
  {Furusaki}},\ }\href {https://doi.org/10.1103/PhysRevLett.88.056402}
  {\bibfield  {journal} {\bibinfo  {journal} {Phys. Rev. Lett.}\ }\textbf
  {\bibinfo {volume} {88}},\ \bibinfo {pages} {056402} (\bibinfo {year}
  {2002})}\BibitemShut {NoStop}%
\bibitem [{\citenamefont {M\'enard}\ and\ \citenamefont
  {Bourbonnais}(2011)}]{Menard2011}%
  \BibitemOpen
  \bibfield  {author} {\bibinfo {author} {\bibfnamefont {M.}~\bibnamefont
  {M\'enard}}\ and\ \bibinfo {author} {\bibfnamefont {C.}~\bibnamefont
  {Bourbonnais}},\ }\href {https://doi.org/10.1103/PhysRevB.83.075111}
  {\bibfield  {journal} {\bibinfo  {journal} {Phys. Rev. B}\ }\textbf {\bibinfo
  {volume} {83}},\ \bibinfo {pages} {075111} (\bibinfo {year}
  {2011})}\BibitemShut {NoStop}%
\bibitem [{\citenamefont {Ejima}\ and\ \citenamefont
  {Nishimoto}(2007)}]{Ejima2007}%
  \BibitemOpen
  \bibfield  {author} {\bibinfo {author} {\bibfnamefont {S.}~\bibnamefont
  {Ejima}}\ and\ \bibinfo {author} {\bibfnamefont {S.}~\bibnamefont
  {Nishimoto}},\ }\href {https://doi.org/10.1103/PhysRevLett.99.216403}
  {\bibfield  {journal} {\bibinfo  {journal} {Phys. Rev. Lett.}\ }\textbf
  {\bibinfo {volume} {99}},\ \bibinfo {pages} {216403} (\bibinfo {year}
  {2007})}\BibitemShut {NoStop}%
\bibitem [{\citenamefont {Iemini}\ \emph {et~al.}(2015)\citenamefont {Iemini},
  \citenamefont {Maciel},\ and\ \citenamefont {Vianna}}]{Iemini2015}%
  \BibitemOpen
  \bibfield  {author} {\bibinfo {author} {\bibfnamefont {F.}~\bibnamefont
  {Iemini}}, \bibinfo {author} {\bibfnamefont {T.~O.}\ \bibnamefont {Maciel}},\
  and\ \bibinfo {author} {\bibfnamefont {R.~O.}\ \bibnamefont {Vianna}},\
  }\href {https://doi.org/10.1103/PhysRevB.92.075423} {\bibfield  {journal}
  {\bibinfo  {journal} {Phys. Rev. B}\ }\textbf {\bibinfo {volume} {92}},\
  \bibinfo {pages} {075423} (\bibinfo {year} {2015})}\BibitemShut {NoStop}%
\bibitem [{\citenamefont {Spalding}\ \emph {et~al.}(2019)\citenamefont
  {Spalding}, \citenamefont {Tsai},\ and\ \citenamefont
  {Campbell}}]{Spalding2019}%
  \BibitemOpen
  \bibfield  {author} {\bibinfo {author} {\bibfnamefont {J.}~\bibnamefont
  {Spalding}}, \bibinfo {author} {\bibfnamefont {S.-W.}\ \bibnamefont {Tsai}},\
  and\ \bibinfo {author} {\bibfnamefont {D.~K.}\ \bibnamefont {Campbell}},\
  }\href {https://doi.org/10.1103/PhysRevB.99.195445} {\bibfield  {journal}
  {\bibinfo  {journal} {Phys. Rev. B}\ }\textbf {\bibinfo {volume} {99}},\
  \bibinfo {pages} {195445} (\bibinfo {year} {2019})}\BibitemShut {NoStop}%
\bibitem [{\citenamefont {Maciejko}(2012)}]{Maciejko2012}%
  \BibitemOpen
  \bibfield  {author} {\bibinfo {author} {\bibfnamefont {J.}~\bibnamefont
  {Maciejko}},\ }\href {https://doi.org/10.1103/PhysRevB.85.245108} {\bibfield
  {journal} {\bibinfo  {journal} {Phys. Rev. B}\ }\textbf {\bibinfo {volume}
  {85}},\ \bibinfo {pages} {245108} (\bibinfo {year} {2012})}\BibitemShut
  {NoStop}%
\bibitem [{\citenamefont {Altshuler}\ \emph {et~al.}(2013)\citenamefont
  {Altshuler}, \citenamefont {Aleiner},\ and\ \citenamefont
  {Yudson}}]{Altshuler2013}%
  \BibitemOpen
  \bibfield  {author} {\bibinfo {author} {\bibfnamefont {B.~L.}\ \bibnamefont
  {Altshuler}}, \bibinfo {author} {\bibfnamefont {I.~L.}\ \bibnamefont
  {Aleiner}},\ and\ \bibinfo {author} {\bibfnamefont {V.~I.}\ \bibnamefont
  {Yudson}},\ }\href {https://doi.org/10.1103/PhysRevLett.111.086401}
  {\bibfield  {journal} {\bibinfo  {journal} {Phys. Rev. Lett.}\ }\textbf
  {\bibinfo {volume} {111}},\ \bibinfo {pages} {086401} (\bibinfo {year}
  {2013})}\BibitemShut {NoStop}%
\bibitem [{\citenamefont {Yevtushenko}\ \emph {et~al.}(2015)\citenamefont
  {Yevtushenko}, \citenamefont {Wugalter}, \citenamefont {Yudson},\ and\
  \citenamefont {Altshuler}}]{Yevtushenko2015}%
  \BibitemOpen
  \bibfield  {author} {\bibinfo {author} {\bibfnamefont {O.~M.}\ \bibnamefont
  {Yevtushenko}}, \bibinfo {author} {\bibfnamefont {A.}~\bibnamefont
  {Wugalter}}, \bibinfo {author} {\bibfnamefont {V.~I.}\ \bibnamefont
  {Yudson}},\ and\ \bibinfo {author} {\bibfnamefont {B.~L.}\ \bibnamefont
  {Altshuler}},\ }\href {https://doi.org/10.1209/0295-5075/112/57003}
  {\bibfield  {journal} {\bibinfo  {journal} {Europhysics Letters}\ }\textbf
  {\bibinfo {volume} {112}},\ \bibinfo {pages} {57003} (\bibinfo {year}
  {2015})}\BibitemShut {NoStop}%
\bibitem [{\citenamefont {Tsvelik}\ and\ \citenamefont
  {Yevtushenko}(2019)}]{Tsvelik2019}%
  \BibitemOpen
  \bibfield  {author} {\bibinfo {author} {\bibfnamefont {A.~M.}\ \bibnamefont
  {Tsvelik}}\ and\ \bibinfo {author} {\bibfnamefont {O.~M.}\ \bibnamefont
  {Yevtushenko}},\ }\href {https://doi.org/10.1103/PhysRevB.100.165110}
  {\bibfield  {journal} {\bibinfo  {journal} {Phys. Rev. B}\ }\textbf {\bibinfo
  {volume} {100}},\ \bibinfo {pages} {165110} (\bibinfo {year}
  {2019})}\BibitemShut {NoStop}%
\bibitem [{\citenamefont {Jordan}\ and\ \citenamefont
  {Wigner}(1928)}]{Jordan1928}%
  \BibitemOpen
  \bibfield  {author} {\bibinfo {author} {\bibfnamefont {P.}~\bibnamefont
  {Jordan}}\ and\ \bibinfo {author} {\bibfnamefont {E.}~\bibnamefont
  {Wigner}},\ }\href {https://doi.org/10.1007/BF01331938} {\bibfield  {journal}
  {\bibinfo  {journal} {Zeitschrift f{\"u}r Physik}\ }\textbf {\bibinfo
  {volume} {47}},\ \bibinfo {pages} {631} (\bibinfo {year} {1928})}\BibitemShut
  {NoStop}%
\bibitem [{\citenamefont {von Delft}\ and\ \citenamefont
  {Schoeller}(1998)}]{Delft1998}%
  \BibitemOpen
  \bibfield  {author} {\bibinfo {author} {\bibfnamefont {J.}~\bibnamefont {von
  Delft}}\ and\ \bibinfo {author} {\bibfnamefont {H.}~\bibnamefont
  {Schoeller}},\ }\href
  {https://doi.org/https://doi.org/10.1002/andp.19985100401} {\bibfield
  {journal} {\bibinfo  {journal} {Annalen der Physik}\ }\textbf {\bibinfo
  {volume} {510}},\ \bibinfo {pages} {225} (\bibinfo {year}
  {1998})}\BibitemShut {NoStop}%
\bibitem [{\citenamefont {Voit}(1995)}]{Voit1995}%
  \BibitemOpen
  \bibfield  {author} {\bibinfo {author} {\bibfnamefont {J.}~\bibnamefont
  {Voit}},\ }\href {https://doi.org/10.1088/0034-4885/58/9/002} {\bibfield
  {journal} {\bibinfo  {journal} {Reports on Progress in Physics}\ }\textbf
  {\bibinfo {volume} {58}},\ \bibinfo {pages} {977} (\bibinfo {year}
  {1995})}\BibitemShut {NoStop}%
\bibitem [{\citenamefont {Fradkin}(2013)}]{Fradkin2013}%
  \BibitemOpen
  \bibfield  {author} {\bibinfo {author} {\bibfnamefont {E.}~\bibnamefont
  {Fradkin}},\ }\href@noop {} {\emph {\bibinfo {title} {{Field Theories of
  Condensed Matter Physics}}}}\ (\bibinfo  {publisher} {Cambridge University
  Press},\ \bibinfo {year} {2013})\BibitemShut {NoStop}%
\bibitem [{\citenamefont {Giamarchi}(2003)}]{Giamarchi2003}%
  \BibitemOpen
  \bibfield  {author} {\bibinfo {author} {\bibfnamefont {T.}~\bibnamefont
  {Giamarchi}},\ }\href@noop {} {\emph {\bibinfo {title} {Quantum Physics in
  One Dimension}}}\ (\bibinfo  {publisher} {Oxford University Press},\ \bibinfo
  {year} {2003})\BibitemShut {NoStop}%
\bibitem [{\citenamefont {Penc}\ and\ \citenamefont
  {S\'olyom}(1993)}]{Penc1993}%
  \BibitemOpen
  \bibfield  {author} {\bibinfo {author} {\bibfnamefont {K.}~\bibnamefont
  {Penc}}\ and\ \bibinfo {author} {\bibfnamefont {J.}~\bibnamefont
  {S\'olyom}},\ }\href {https://doi.org/10.1103/PhysRevB.47.6273} {\bibfield
  {journal} {\bibinfo  {journal} {Phys. Rev. B}\ }\textbf {\bibinfo {volume}
  {47}},\ \bibinfo {pages} {6273} (\bibinfo {year} {1993})}\BibitemShut
  {NoStop}%
\bibitem [{\citenamefont {Zhao}\ and\ \citenamefont {Liu}(2008)}]{Zhao2008}%
  \BibitemOpen
  \bibfield  {author} {\bibinfo {author} {\bibfnamefont {E.}~\bibnamefont
  {Zhao}}\ and\ \bibinfo {author} {\bibfnamefont {W.~V.}\ \bibnamefont {Liu}},\
  }\href {https://doi.org/10.1103/PhysRevA.78.063605} {\bibfield  {journal}
  {\bibinfo  {journal} {Phys. Rev. A}\ }\textbf {\bibinfo {volume} {78}},\
  \bibinfo {pages} {063605} (\bibinfo {year} {2008})}\BibitemShut {NoStop}%
\bibitem [{\citenamefont {Capponi}\ \emph {et~al.}(2000)\citenamefont
  {Capponi}, \citenamefont {Poilblanc},\ and\ \citenamefont
  {Giamarchi}}]{Capponi2000}%
  \BibitemOpen
  \bibfield  {author} {\bibinfo {author} {\bibfnamefont {S.}~\bibnamefont
  {Capponi}}, \bibinfo {author} {\bibfnamefont {D.}~\bibnamefont {Poilblanc}},\
  and\ \bibinfo {author} {\bibfnamefont {T.}~\bibnamefont {Giamarchi}},\ }\href
  {https://doi.org/10.1103/PhysRevB.61.13410} {\bibfield  {journal} {\bibinfo
  {journal} {Phys. Rev. B}\ }\textbf {\bibinfo {volume} {61}},\ \bibinfo
  {pages} {13410} (\bibinfo {year} {2000})}\BibitemShut {NoStop}%
\bibitem [{\citenamefont {Affleck}\ and\ \citenamefont
  {Giuliano}(2013)}]{Affleck2013}%
  \BibitemOpen
  \bibfield  {author} {\bibinfo {author} {\bibfnamefont {I.}~\bibnamefont
  {Affleck}}\ and\ \bibinfo {author} {\bibfnamefont {D.}~\bibnamefont
  {Giuliano}},\ }\href {https://doi.org/10.1088/1742-5468/2013/06/p06011}
  {\bibfield  {journal} {\bibinfo  {journal} {Journal of Statistical Mechanics:
  Theory and Experiment}\ }\textbf {\bibinfo {volume} {2013}},\ \bibinfo
  {pages} {P06011} (\bibinfo {year} {2013})}\BibitemShut {NoStop}%
\bibitem [{\citenamefont {Jones}(2017)}]{Jones2017PHD}%
  \BibitemOpen
  \bibfield  {author} {\bibinfo {author} {\bibfnamefont {M.}~\bibnamefont
  {Jones}},\ }\emph {\bibinfo {title} {Disorder in Multi-Channel Luttinger
  Liquids}},\ \href@noop {} {Ph.D. thesis},\ \bibinfo  {school} {The University
  of Birmingham} (\bibinfo {year} {2017})\BibitemShut {NoStop}%
\bibitem [{Sup()}]{Suppl}%
  \BibitemOpen
  \href@noop {} {}\bibinfo {note} {See Supplemental Material on page
  \pageref{sec:supp} for details regarding the computation of correlation
  functions using the Luttinger matrix formalism.}\BibitemShut {Stop}%
\bibitem [{\citenamefont {Fulde}\ and\ \citenamefont
  {Ferrell}(1964)}]{Fulde1964}%
  \BibitemOpen
  \bibfield  {author} {\bibinfo {author} {\bibfnamefont {P.}~\bibnamefont
  {Fulde}}\ and\ \bibinfo {author} {\bibfnamefont {R.~A.}\ \bibnamefont
  {Ferrell}},\ }\href {https://doi.org/10.1103/PhysRev.135.A550} {\bibfield
  {journal} {\bibinfo  {journal} {Phys. Rev.}\ }\textbf {\bibinfo {volume}
  {135}},\ \bibinfo {pages} {A550} (\bibinfo {year} {1964})}\BibitemShut
  {NoStop}%
\bibitem [{\citenamefont {Larkin}\ and\ \citenamefont
  {Ovchinnikov}(1964)}]{Larkin1964}%
  \BibitemOpen
  \bibfield  {author} {\bibinfo {author} {\bibfnamefont {A.~I.}\ \bibnamefont
  {Larkin}}\ and\ \bibinfo {author} {\bibfnamefont {Y.~N.}\ \bibnamefont
  {Ovchinnikov}},\ }\href {https://doi.org/https://www.osti.gov/biblio/4653415}
  {\bibfield  {journal} {\bibinfo  {journal} {Zh. Eksperim. i Teor. Fiz.}\
  }\textbf {\bibinfo {volume} {47}},\ \bibinfo {pages} {1136} (\bibinfo {year}
  {1964})}\BibitemShut {NoStop}%
\bibitem [{\citenamefont {Yang}(2001)}]{Yang2001}%
  \BibitemOpen
  \bibfield  {author} {\bibinfo {author} {\bibfnamefont {K.}~\bibnamefont
  {Yang}},\ }\href {https://doi.org/10.1103/PhysRevB.63.140511} {\bibfield
  {journal} {\bibinfo  {journal} {Phys. Rev. B}\ }\textbf {\bibinfo {volume}
  {63}},\ \bibinfo {pages} {140511} (\bibinfo {year} {2001})}\BibitemShut
  {NoStop}%
\bibitem [{\citenamefont {Feiguin}\ and\ \citenamefont
  {Huse}(2009)}]{Feigun2009}%
  \BibitemOpen
  \bibfield  {author} {\bibinfo {author} {\bibfnamefont {A.~E.}\ \bibnamefont
  {Feiguin}}\ and\ \bibinfo {author} {\bibfnamefont {D.~A.}\ \bibnamefont
  {Huse}},\ }\href {https://doi.org/10.1103/PhysRevB.79.100507} {\bibfield
  {journal} {\bibinfo  {journal} {Phys. Rev. B}\ }\textbf {\bibinfo {volume}
  {79}},\ \bibinfo {pages} {100507} (\bibinfo {year} {2009})}\BibitemShut
  {NoStop}%
\bibitem [{\citenamefont {Kitazawa}(2003)}]{Kitazawa2003}%
  \BibitemOpen
  \bibfield  {author} {\bibinfo {author} {\bibfnamefont {A.}~\bibnamefont
  {Kitazawa}},\ }\href {https://doi.org/10.1088/0953-8984/15/17/313} {\bibfield
   {journal} {\bibinfo  {journal} {Journal of Physics: Condensed Matter}\
  }\textbf {\bibinfo {volume} {15}},\ \bibinfo {pages} {2587} (\bibinfo {year}
  {2003})}\BibitemShut {NoStop}%
\bibitem [{\citenamefont {Wentzel}(1951)}]{Wentzel1951}%
  \BibitemOpen
  \bibfield  {author} {\bibinfo {author} {\bibfnamefont {G.}~\bibnamefont
  {Wentzel}},\ }\href {https://doi.org/10.1103/PhysRev.83.168} {\bibfield
  {journal} {\bibinfo  {journal} {Phys. Rev.}\ }\textbf {\bibinfo {volume}
  {83}},\ \bibinfo {pages} {168} (\bibinfo {year} {1951})}\BibitemShut
  {NoStop}%
\bibitem [{\citenamefont {Bardeen}(1951)}]{Bardeen1951}%
  \BibitemOpen
  \bibfield  {author} {\bibinfo {author} {\bibfnamefont {J.}~\bibnamefont
  {Bardeen}},\ }\href {https://doi.org/10.1103/RevModPhys.23.261} {\bibfield
  {journal} {\bibinfo  {journal} {Rev. Mod. Phys.}\ }\textbf {\bibinfo {volume}
  {23}},\ \bibinfo {pages} {261} (\bibinfo {year} {1951})}\BibitemShut
  {NoStop}%
\bibitem [{\citenamefont {Ogata}\ \emph {et~al.}(1991)\citenamefont {Ogata},
  \citenamefont {Luchini}, \citenamefont {Sorella},\ and\ \citenamefont
  {Assaad}}]{Ogata1991_2}%
  \BibitemOpen
  \bibfield  {author} {\bibinfo {author} {\bibfnamefont {M.}~\bibnamefont
  {Ogata}}, \bibinfo {author} {\bibfnamefont {M.~U.}\ \bibnamefont {Luchini}},
  \bibinfo {author} {\bibfnamefont {S.}~\bibnamefont {Sorella}},\ and\ \bibinfo
  {author} {\bibfnamefont {F.~F.}\ \bibnamefont {Assaad}},\ }\href
  {https://doi.org/10.1103/PhysRevLett.66.2388} {\bibfield  {journal} {\bibinfo
   {journal} {Phys. Rev. Lett.}\ }\textbf {\bibinfo {volume} {66}},\ \bibinfo
  {pages} {2388} (\bibinfo {year} {1991})}\BibitemShut {NoStop}%
\bibitem [{\citenamefont {Loss}\ and\ \citenamefont {Martin}(1994)}]{Loss1994}%
  \BibitemOpen
  \bibfield  {author} {\bibinfo {author} {\bibfnamefont {D.}~\bibnamefont
  {Loss}}\ and\ \bibinfo {author} {\bibfnamefont {T.}~\bibnamefont {Martin}},\
  }\href {https://doi.org/10.1103/PhysRevB.50.12160} {\bibfield  {journal}
  {\bibinfo  {journal} {Phys. Rev. B}\ }\textbf {\bibinfo {volume} {50}},\
  \bibinfo {pages} {12160} (\bibinfo {year} {1994})}\BibitemShut {NoStop}%
\bibitem [{\citenamefont {Hsu}\ \emph {et~al.}(2017)\citenamefont {Hsu},
  \citenamefont {Stano}, \citenamefont {Klinovaja},\ and\ \citenamefont
  {Loss}}]{Hsu2017}%
  \BibitemOpen
  \bibfield  {author} {\bibinfo {author} {\bibfnamefont {C.-H.}\ \bibnamefont
  {Hsu}}, \bibinfo {author} {\bibfnamefont {P.}~\bibnamefont {Stano}}, \bibinfo
  {author} {\bibfnamefont {J.}~\bibnamefont {Klinovaja}},\ and\ \bibinfo
  {author} {\bibfnamefont {D.}~\bibnamefont {Loss}},\ }\href
  {https://doi.org/10.1103/PhysRevB.96.081405} {\bibfield  {journal} {\bibinfo
  {journal} {Phys. Rev. B}\ }\textbf {\bibinfo {volume} {96}},\ \bibinfo
  {pages} {081405} (\bibinfo {year} {2017})}\BibitemShut {NoStop}%
\end{thebibliography}%


%

\clearpage
\onecolumngrid
\allowdisplaybreaks

\renewcommand{\thefigure}{S\arabic{figure}}
\renewcommand{\theHfigure}{S\arabic{figure}}
\setcounter{figure}{0}  


\renewcommand{\thetable}{S\Roman{table}}
\setcounter{table}{0}  

\renewcommand{\theequation}{S\arabic{equation}}
\setcounter{equation}{0}  

\renewcommand{\thesection}{S\arabic{section}}
\renewcommand{\thesubsection}{\thesection.\arabic{subsection}}
\renewcommand{\thesubsubsection}{\thesubsection.\arabic{subsubsection}}
\makeatletter
\renewcommand{\p@subsection}{}
\renewcommand{\p@subsubsection}{}
\makeatother

\phantomsection \label{sec:supp}

\section{Calculation of correlation functions in multi-channel Luttinger liquids} 
This Supplemental material is meant to elucidate the calculation of correlation functions for multi-channel Luttinger liquids using the Luttinger matrix formalism. Starting from the action describing a general $n$-channel Luttinger liquid
\begin{equation}
    S[\boldsymbol{\phi}, \boldsymbol{\theta}] = \frac{1}{2\pi} \int \mathrm{d} x \mathrm{d} \tau \begin{pmatrix}
        \boldsymbol{\phi}^{\mathrm{T}} & \boldsymbol{\theta}^{\mathrm{T}}
    \end{pmatrix}
    \left[
        \begin{pmatrix}
            0 & \mathbb{I}_n \\
            \mathbb{I}_n & 0
        \end{pmatrix}i\partial_\tau + \begin{pmatrix}
            V_\phi & 0 \\
            0 & V_\theta
        \end{pmatrix}\partial_x
    \right] \partial_x\begin{pmatrix}
        \boldsymbol{\phi} \\ \boldsymbol{\theta},
    \end{pmatrix}, 
    \label{action_fields_Sup_Mat}
\end{equation}
where the $n\times n$ interaction matrices $V_\phi$ and $V_\theta$ are not diagonal, we will demonstrate how to calculate correlation functions in the form
\begin{equation}
    I = \langle  \mathrm{exp}(i\sum_i \boldsymbol{A}_i^{\mathrm{T}} \boldsymbol{\phi}(r_i) + \boldsymbol{B}_i^{\mathrm{T}} \boldsymbol{\theta}(r_i)) \rangle.
    \label{I_integra_Sup_Mat}
\end{equation}
To serve as an example, consider the correlation function for the spin-polarized triplet state, $R_{\mathrm{TS}^\uparrow}$. In terms of the bosonic fields, $R_{\mathrm{TS}^\uparrow}(x, \tau=0)$ is given by
\begin{equation}
    R_{\mathrm{TS}^\uparrow}(x) = \frac{1}{(2\pi\alpha)^2} \big\langle \mathrm{e}^{\sqrt{2}i(\theta_\rho(x)-\theta_\rho(0)) -\sqrt{2}i(\theta_\sigma(x)-\theta_\sigma(0))}\rangle. 
    \label{R_TS}
\end{equation}
Calculating $R_{\mathrm{TS}^\uparrow}$ is thus equivalent to calculating $I$ with spatial coordinates $x_1 =x$, $x_2 = 0$, $\tau_1=\tau_2=0$, and coefficients $B_1^1 =\sqrt{2},  B_2^1 =-\sqrt{2}, B_1^2 =-\sqrt{2},  B_2^2 =\sqrt{2}$. The other coefficients in $\boldsymbol{B}$ as well as all coefficients in $\boldsymbol{A}$ are zero.

As described in the main text, one can find a matrix $M$ relating the original fields $\boldsymbol{\phi}$ and $\boldsymbol{\theta}$ to new fields $\tilde{\boldsymbol{\phi}}$ and $\tilde{\boldsymbol{\theta}}$ which diagonalize equation \eqref{action_fields_Sup_Mat}. By employing the following transformation
\begin{subequations}
    \label{M_trans}
    \begin{align}
        \boldsymbol{\phi} &= M \tilde{\boldsymbol{\phi}} \\
        \boldsymbol{\theta} &= M^{-\mathrm{T}} \tilde{\boldsymbol{\theta}} \label{M_theta_trans},
    \end{align}
\end{subequations}
and inserting into equation \eqref{action_fields_Sup_Mat}, we obtain
\begin{equation}
    S[\tilde{\boldsymbol{\phi}}, \tilde{\boldsymbol{\theta}}] = \frac{1}{2\pi} \int  \mathrm{d} x \mathrm{d} \tau \begin{pmatrix}
        \tilde{\boldsymbol{\phi}}^{\mathrm{T}} & \tilde{\boldsymbol{\theta}}^{\mathrm{T}}
    \end{pmatrix}
    \left[
        \begin{pmatrix}
            0 & \mathbb{I}_n \\
            \mathbb{I}_n & 0
        \end{pmatrix}i\partial_\tau + \begin{pmatrix}
            u & 0 \\
            0 & u
        \end{pmatrix}\partial_x
    \right] \partial_x\begin{pmatrix}
        \tilde{\boldsymbol{\phi}} \\ \tilde{\boldsymbol{\theta}}
    \end{pmatrix},
    \label{action_fields_diag_Sup_Mat}
\end{equation}
where $u$ is a diagonal matrix.

We will now calculate $I$ using equations \eqref{M_trans} and \eqref{action_fields_diag_Sup_Mat}. The calculation follows the single-channel case in \cite{Giamarchi2003} closely, the main point is to show how one accounts for the basis change required when $V_\phi$ and $V_\theta$ are not diagonal. We consider the case where only $\boldsymbol{B}$ contains nonzero values, and insert the aforementioned coefficients associated with $R_{\mathrm{TS}^\uparrow}$ at the end of calculation. This specific configuration of $I$ is thus $I_\theta = \big\langle  \mathrm{e}^{i\sum_i \boldsymbol{B}_i^{\mathrm{T}} \boldsymbol{\theta}(r_i)} \big\rangle$. The calculation of other correlation functions is analogous. The first step of the calculation is to rewrite the fields in their Fourier representation
\begin{equation}
    \sum_i \boldsymbol{B}_i^{\mathrm{T}} \boldsymbol{\theta}(r_i) = \frac{1}{2\beta\Omega} \sum_k((\boldsymbol{B}^*)^{\mathrm{T}}(k)\boldsymbol{\theta}(k) +\mathrm{h.c.}),
\label{fourier_B}
\end{equation}
with $k$ as the shorthand notation for $k=(q,\omega_n)$ where $q$ is a wave vector and $\omega_n$ is a bosonic Matsubara frequency. $\Omega$ and $\beta$ denote the system size and inverse temperature, respectively. We also introduced
\begin{equation}
    \boldsymbol{B}(k) \equiv \sum_i \mathrm{e}^{-ir_ik}\boldsymbol{B}_i.
    \label{shorthand_fourier}
\end{equation}

Since $S$ is not diagonal in the original fields we must rewrite equation \eqref{fourier_B} in terms of $\boldsymbol{\tilde{\theta}}$ by using equation \eqref{M_theta_trans}
\begin{equation}
    \frac{1}{2\beta\Omega} \sum_k \sum_l B^{l*}(k) \theta_l(k) + B^l(k) \theta^*_l(k) = \frac{1}{2\beta\Omega} \sum_k \sum_l \bigg[B^{l*}(k)(k) \sum_m M_{lm}^{-\mathrm{T}}\tilde{\theta}_m(k) + B^l(k) \sum_m M_{lm}^{-\mathrm{T}}\tilde{\theta}^*_m(k)\bigg]
    \label{fourier_B_tilde}. 
\end{equation}
Because $I_\theta$ is independent of $\phi$, one can integrate out the density fields in $S[\tilde{\boldsymbol{\phi}}, \tilde{\boldsymbol{\theta}}]$, akin to the textbook procedure used when considering the one-channel case. The resulting action is 
\begin{equation}
    S[\boldsymbol{\tilde{\theta}}] = \frac{1}{2\pi}\int \mathrm{d} x \mathrm{d} \tau \sum_m \frac{1}{u_m}((\partial_\tau\tilde{\theta}_m)^2 + (u_m\partial_x\tilde{\theta}_m)^2)\label{S_theta},
\end{equation}
where $u_m$ is the $m$'th diagonal element in $u$.
Using the standard formula for integrating a multivariate Gaussian integral, one can integrate over the density sector by using equations \eqref{fourier_B_tilde} and \eqref{S_theta}
\begin{align}
    I_\theta &= \frac{1}{\mathcal{Z}_\theta}\int\mathcal{D} \boldsymbol{\theta}\exp(-\frac{1}{2\pi\beta\Omega}\sum_k \sum_m \bigg[\frac{\tilde{\theta}_m(k)\tilde{\theta}_m^*(k)}{u_m}(\omega_n^2 + u_m^2q^2) -i\pi (
        \tilde{\theta}_m(k) \sum_l B^{l*}(k)M_{lm}^{-\mathrm{T}} + \tilde{\theta}^*_m(k)\sum_{l'} B^{l'}(k)M_{l'm}^{-\mathrm{T}} 
    )\bigg])\\
    &= \exp(
        -\frac{\pi}{2\beta\Omega}\sum_k\sum_m\frac{u_m}{\omega_n^2 + u_mq^2}\sum_{l,l'}B^{l*}(k)M_{lm}^{-\mathrm{T}}B^{l'}(k)M_{l'm}^{-\mathrm{T}} 
    ) \\
    &= \exp(-
    \sum_{l,l'} \sum_mM_{lm}^{-\mathrm{T}} M_{l'm}^{-\mathrm{T}}\frac{\pi}{2\beta\Omega}\sum_{i,j}B_i^lB^{l'}_j \sum_k \mathrm{e}^{i(r_i-r_j)k}\frac{u_m}{\omega_n^2 + u_m^2q^2}),
\end{align}
where we reinserted equation \eqref{shorthand_fourier} in going from the second to third line. $\mathcal{Z}_\theta$ is the partition function for the system after integrating out the density fields. By employing the same trick as in \cite{Giamarchi2003}, adding and subtracting the same quantity in the exponent, we obtain
\begin{align}
    \begin{split}
        I_\theta = &\exp(-\frac{\pi}{2\beta\Omega}
    \sum_{l,l'} \sum_mM_{lm}^{-\mathrm{T}} M_{l'm}^{-\mathrm{T}}\sum_{i,j}B_i^lB^{l'}_j \sum_k (\mathrm{e}^{i(r_i-r_j)k} -1)\frac{u_m}{\omega_n^2 + u_m^2q^2}) \\
    &\exp(-\frac{\pi}{2\beta\Omega}
    \sum_{l,l'} \sum_mM_{lm}^{-\mathrm{T}} M_{l'm}^{-\mathrm{T}}\sum_{i,j}B_i^lB^{l'}_j \sum_k \frac{u_m}{\omega_n^2 + u_m^2q^2}).
\end{split}
\end{align}
The last exponential can be further simplified
\begin{equation}
    \exp(-\frac{\pi}{2\beta\Omega}\sum_{l,l'} \sum_mM_{lm}^{-\mathrm{T}} M_{l'm}^{-\mathrm{T}}\sum_{i,j}B_i^lB^{l'}_j \sum_k \frac{u_m}{\omega_n^2 + u_m^2q^2}) = \exp(-\frac{\pi}{2\beta\Omega}\sum_m \left( \sum_{l} \sum_i M_{lm}^{-\mathrm{T}} B_i^l\right)^2\sum_k \frac{u_m}{\omega_n^2 + u_m^2q^2}).
\end{equation}
Unless the quantity $\sum_{l} \sum_i M_{lm}^{-\mathrm{T}} B_i^l$ is zero, the exponent diverges, thus exponentially suppressing $I_\theta$. This gives rise to the criterion stated in the main text, namely that $\sum_{l} \sum_i M_{lm}^{-\mathrm{T}} B_i^l=0$. Assuming that the criterion is fulfilled, $I_\theta$ the Matsubara sum and the wave vector integral is calculated following \cite{Giamarchi2003} in the zero-temperature limit, and we finally arrive at the general expression
\begin{align}
    I_\theta = \prod_m\biggl(
        \frac{\alpha^2}{(x_i-x_j)^2 +(u_m|\tau_i - \tau_j| + \alpha)^2}
    \biggr)^{
        -\frac{1}{4}\sum_{l,l'}\sum_{i<j} M_{lm}^{-\mathrm{T}} M_{ml'}^{-1} A_i^lA_j^{l'}
    }.
    \label{I_solved}
\end{align}

By inserting the coefficients used in $R_{\mathrm{TS}^\uparrow}$, we obtain 
\begin{equation}
    R_{\mathrm{TS}^\uparrow}(x) = \frac{1}{(2\pi\alpha)^2} \frac{1}{x^{K^{-1}_{11} + K^{-1}_{22} + K^{-1}_{12}}}, 
    \label{R_TS_solved}
\end{equation}
as stated in the main text. Computations of the remaining correlation functions can be carried out in a similar manner.

\end{document}